\documentclass[journal]{IEEEtran}
\usepackage[T1]{fontenc}
\usepackage[utf8]{inputenc}
\usepackage[binary-units=true]{siunitx}
\DeclareSIUnit{\bits}{bits}
\usepackage{dsfont}
\usepackage{amssymb}
\usepackage[pdftex]{graphicx}
\usepackage{subcaption} 
\usepackage{amsmath}
\usepackage{amsthm}
\usepackage{mathtools}
\usepackage{array}
\usepackage[noadjust]{cite}
\usepackage{xcolor}
\usepackage{tikz}
\usepackage{stfloats}
\usepackage[draft]{hyperref}
\usepackage[acronym]{glossaries}
\usepackage{hhline}
\usepackage{multirow}
\usepackage{flushend}

\DeclareFontFamily{OT1}{pzc}{}
\DeclareFontShape{OT1}{pzc}{m}{it}{<-> s * [1.100] pzcmi7t}{}
\DeclareMathAlphabet{\mathscr}{OT1}{pzc}{m}{it}

\usepackage[export]{adjustbox}
\usepackage{pgfplots}
\DeclareUnicodeCharacter{2212}{−}
\usepgfplotslibrary{groupplots,dateplot}
\usetikzlibrary{patterns,shapes.arrows}
\pgfplotsset{compat=newest}

\newenvironment{frcseries}{\fontfamily{frc}\selectfont}{}

\usepackage{amsfonts} 

\newcommand{\shortminus}{\scalebox{0.75}[1.0]{\( - \)}}

\usetikzlibrary{decorations.pathreplacing,calc}

\newcommand{\addstretch}[1]{\addtolength{#1}{\fill}}

\makeatletter
\let\MYcaption\@makecaption
\makeatother

\usepackage[font=footnotesize]{subcaption}

\makeatletter
\let\@makecaption\MYcaption
\makeatother

\newacronym{3GPP}{3GPP}{3rd Generation Partnership Project}
\newacronym{ACM}{ACM}{adaptive coding and modulation}
\newacronym{ADC}{ADC}{analog-to-digital conversion}
\newacronym{AGC}{AGC}{automatic gain control}
\newacronym{AWGN}{AWGN}{additive white Gaussian noise}
\newacronym{BER}{BER}{bit error rate}
\newacronym{BS}{BS}{base station}
\newacronym{BLER}{BLER}{block error rate}
\newacronym{BCE}{BCE}{binary cross-entropy}
\newacronym{BMD}{BMD}{bit-metric decoding}
\newacronym{BP}{BP}{backpropagation}
\newacronym{BPTT}{BPTT}{backpropagation through time}
\newacronym{CE}{CE}{cross-entropy}
\newacronym{CFO}{CFO}{carrier frequency offset}
\newacronym{CNN}{CNN}{convolutional neural network}
\newacronym{CSI}{CSI}{channel state information}
\newacronym{DAC}{DAC}{digital-to-analog conversion}
\newacronym{DL}{DL}{deep learning}
\newacronym{DFT}{DFT}{discrete Fourier transform}
\newacronym{ELU}{ELU}{exponential linear unit}
\newacronym{FFT}{FFT}{fast Fourier transform}
\newacronym{GAN}{GAN}{generative adversarial network}
\newacronym{GRU}{GRU}{gated recurrent unit}
\newacronym{iid}{i.i.d.\@}{independent and identically distributed}
\newacronym{IFFT}{IFFT}{inverse fast Fourier transform}
\newacronym{KL}{KL}{Kullback-Leibler}
\newacronym{LLR}{LLR}{log likelihood ratio}
\newacronym{LSTM}{LSTM}{long short-term memory}
\newacronym{LDPC}{LDPC}{low-density parity-check}
\newacronym{LMMSE}{LMMSE}{linear minimum mean squared error}
\newacronym{MDP}{MDP}{Markov decision process}
\newacronym{ML}{ML}{machine learning}
\newacronym{MLP}{MLP}{multilayer perceptron}
\newacronym{MIMO}{MIMO}{multiple-input multiple-output}
\newacronym{MU-MIMO}{MU-MIMO}{multi-user multiple-input multiple-output}
\newacronym{MU}{MU}{multi-user}
\newacronym{MSE}{MSE}{mean squared error}
\newacronym{NN}{NN}{neural network}
\newacronym{NR}{NR}{new radio}
\newacronym{NLOS}{NLOS}{non-line of sight}
\newacronym{NIRE}{NIRE}{nearest interpolated resource element}
\newacronym{OFDM}{OFDM}{orthogonal frequency-division multiplexing}
\newacronym{pdf}{pdf}{probability density function}
\newacronym{pmf}{pmf}{probability mass function}
\newacronym{QPSK}{QPSK}{quadrature phase-shift keying}
\newacronym{QAM}{QAM}{quadrature amplitude modulation}
\newacronym{PSNR}{PSNR}{Peak Signal to Noise Ratio}
\newacronym{RBF}{RBF}{Rayleigh block-fading}
\newacronym{RB}{RB}{resource block}
\newacronym{RE}{RE}{resource element}
\newacronym{RG}{RG}{resource grid\newacronym{RE}{RE}{resource element}}
\newacronym{ReLU}{ReLU}{rectified linear unit}
\newacronym{RTN}{RTN}{radio transformer network}
\newacronym{RL}{RL}{reinforcement learning}
\newacronym{RNN}{RNN}{recurrent neural network}
\newacronym{SFO}{SFO}{sampling frequency offset}
\newacronym{SER}{SER}{symbol error rate}
\newacronym{SNR}{SNR}{signal-to-noise ratio}
\newacronym{SINR}{SINR}{signal-to-interference-plus-noise ratio}
\newacronym{SGD}{SGD}{stochastic gradient descent}
\newacronym{SISO}{SISO}{single-input single-output}
\newacronym{SIMO}{SIMO}{single-input multiple-output}
\newacronym{SU}{SU}{single-user}
\newacronym{TDD}{TDD}{time-division duplexing}
\newacronym{TR}{TR}{technical report}
\newacronym{UE}{UE}{user equipment}
\newacronym{UMi}{UMi}{urban microcell}
\newacronym{ULA}{ULA}{uniform linear array}
\newacronym{wrt}{w.r.t.\@}{with respect to}

\renewcommand{\vec}[1]{\mathbf{#1}}
\newcommand{\vecs}[1]{\boldsymbol{#1}}

\newcommand{\gv}{\vec{g}}
\newcommand{\hv}{\vec{h}}

\newcommand{\nv}{\vec{n}}

\newcommand{\qv}{\vec{q}}
\newcommand{\rv}{\vec{r}}
\newcommand{\sv}{\vec{s}}
\newcommand{\tv}{\vec{t}}
\newcommand{\uv}{\vec{u}}
\newcommand{\vv}{\vec{v}}
\newcommand{\wv}{\vec{w}}
\newcommand{\xv}{\vec{x}}
\newcommand{\yv}{\vec{y}}
\newcommand{\zv}{\vec{z}}
\newcommand{\zerov}{\vec{0}}

\newcommand{\thetav}{\vecs{\theta}}

\newcommand{\Dm}{\vec{D}}
\newcommand{\Em}{\vec{E}}

\newcommand{\Gm}{\vec{G}}
\newcommand{\Hm}{\vec{H}}
\newcommand{\Id}{\vec{I}}

\newcommand{\Mm}{\vec{M}}
\newcommand{\Nm}{\vec{N}}

\newcommand{\Rm}{\vec{R}}
\newcommand{\Sm}{\vec{S}}
\newcommand{\Tm}{\vec{T}}
\newcommand{\Um}{\vec{U}}
\newcommand{\Vm}{\vec{V}}
\newcommand{\Wm}{\vec{W}}
\newcommand{\Xm}{\vec{X}}
\newcommand{\Ym}{\vec{Y}}

\newcommand{\Sigmam}{\vecs{\Sigma}}
\newcommand{\Omegam}{\vecs{\Omega}}

\newcommand{\Psim}{\vecs{\Psi}}

\newcommand{\Cc}{{\cal C}}

\newcommand{\Nc}{{\cal N}}

\newcommand{\Pc}{{\cal P}}

\newcommand{\CC}{\mathbb{C}}

\newcommand{\RR}{\mathbb{R}}

\newcommand{\htp}{^{\mathsf{H}}}
\newcommand{\tp}{^{\mathsf{T}}}

\newcommand{\LB}{\left(}
\newcommand{\RB}{\right)}
\newcommand{\LP}{\left\{}
\newcommand{\RP}{\right\}}
\newcommand{\LSB}{\left[}
\newcommand{\RSB}{\right]}

\renewcommand{\ln}[1]{\mathop{\mathrm{ln}}\LB #1\RB}
\renewcommand{\log}[1]{\mathop{\mathrm{log}}\LB #1\RB}
\renewcommand{\exp}[1]{\mathop{\mathrm{exp}}\LB #1\RB}

\newcommand{\EE}{{\mathbb{E}}}

\newlength{\dhatheight}

\newcommand\abs[1]{\mathopen|#1\mathclose|}

\begin{document}
\title{Machine Learning for MU-MIMO \\ Receive Processing in OFDM Systems}

\author{Mathieu~Goutay,~
\IEEEmembership{Student Member,~IEEE},
        Fayçal~Ait~Aoudia,
       	~\IEEEmembership{Member,~IEEE},
        ~Jakob~Hoydis,
        ~\IEEEmembership{Senior~Member,~IEEE},
        and Jean-Marie~Gorce,
        ~\IEEEmembership{Senior~Member,~IEEE}
\thanks{F. Ait Aoudia and J. Hoydis are with Nokia Bell Labs, Paris-Saclay, 91620 Nozay, France (e-mail: \{faycal.ait\_aoudia, jakob.hoydis\}@nokia-bell-labs.com). }
\thanks{J.-M. Gorce is with Univ. Lyon, INSA Lyon, Inria, CITI, 69100 Villeurbanne, France (e-mail: jean-marie.gorce@insa-lyon.fr).}
\thanks{M. Goutay is with both institutions (e-mail: mathieu.goutay@nokia.com).}}

\maketitle

\begin{abstract}

\Gls{ML} starts to be widely used to enhance the performance of \gls{MU-MIMO} receivers. 
However, it is still unclear if such methods are truly competitive with respect to conventional methods in realistic scenarios and under practical constraints. 
In addition to enabling accurate signal reconstruction on realistic channel models, \gls{MU-MIMO} receive algorithms must allow for easy adaptation to a varying number of users without the need for retraining. 
In contrast to existing work, we propose an \gls{ML}-enhanced \gls{MU-MIMO} receiver that builds on top of a conventional \gls{LMMSE} architecture. 
It preserves the interpretability and scalability of the \gls{LMMSE} receiver, while improving its accuracy in two ways. 
First, \glspl{CNN} are used to compute an approximation of the second-order statistics of the channel estimation error which are required for accurate equalization. 
Second, a \gls{CNN}-based demapper jointly processes a large number of \gls{OFDM} symbols and subcarriers, which allows it to compute better \glspl{LLR} by compensating for channel aging.
The resulting architecture can be used in the up-~and downlink and is trained in an end-to-end manner, removing the need for hard-to-get perfect \gls{CSI} during the training phase. 
Simulation results demonstrate consistent performance improvements over the baseline which are especially pronounced in high mobility scenarios.

\begin{IEEEkeywords}
Multi-user MIMO detection, OFDM, channel estimation, deep learning, neural networks
\end{IEEEkeywords}

\end{abstract}

\glsresetall

\section{Introduction} 
\label{sec:introduction}

\Gls{ML} has transformed many research areas in the past decade but only started to revolutionize wireless communications in recent years with applications to end-to-end learning \cite{8054694}, channel coding \cite{7852251}, resource management \cite{8227766}, and many others \cite{8839651}. 
In particular, the application of \gls{ML} to the physical layer has led to promising results on realistic channels \cite{8335670} and this technology is now perceived as a major enabler of future generations of wireless networks \cite{9078454}.
Another key technique is the use of \glsdesc{MU}\glsunset{MU} \glsdesc{MIMO}\glsunset{MIMO} (MU-MIMO) systems, where spatial multiplexing is leveraged to increase both the channel capacity and the number of users that can be served simultaneously~\cite{massivemimobook}.
One of the main challenges related to the deployment of such systems is the complexity of the receive processing, which grows with the number of antennas and users. 
Traditional non-linear \gls{MIMO} detection methods offer strong performance but quickly become impractical in large systems. 
For example, maximum a posteriori detection is optimal but known to be NP-hard, and sphere decoders have exponential worst-case complexity \cite{7244171}.
The conventional solution to tackle this problem uses linear detectors that are computationally tractable, but suffers from performance degradation on ill-conditioned channels.
Therefore, the development of new receivers that combine low complexity and high performance is crucial to unlock the full potential of \gls{MU}-\gls{MIMO}.

\begin{figure}[h]
    \centering
    \includegraphics[width=0.49\textwidth]{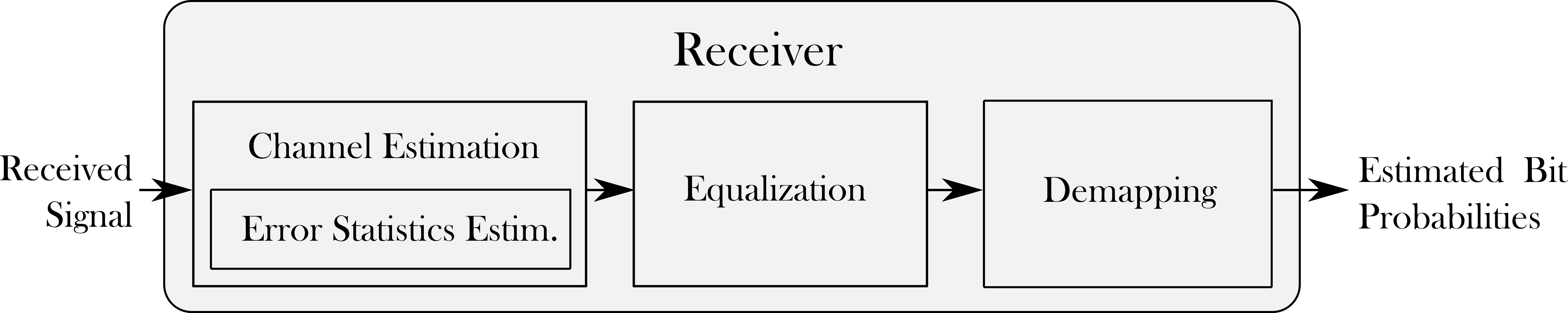}
    \caption{Traditional receiver architecture.}
    \label{fig:trad_receiver}
\end{figure}

An attractive research direction is to apply \gls{ML} techniques to improve \gls{MIMO} reception.
Essentially, two distinct approaches are emerging.
The first one is to consider a single processing block in the typical receiver architecture depicted in Fig.~\ref{fig:trad_receiver} and to augment it with \gls{ML}.
The second approach is to replace multiple blocks with a single \gls{ML} component, such as a \gls{NN}.
The former has the advantages of remaining interpretable and scalable to any number of users \cite{pratik2020remimo}, but is not trained to optimize the end-to-end receiver performance and requires ground truth measurements that are not available in practice.
The latter has shown promising results \cite{korpi2020deeprx}, but is very computationally expensive and acts as a black-box that cannot be adapted to changing number of users or transmit antennas without full retraining.

In this paper, we introduce a new hybrid strategy, where multiple \gls{ML} components are jointly trained to enhance a conventional \gls{MU}-\gls{MIMO} receiver architecture.
The goal is to combine the interpretability of the first approach, the efficiency of the second one, and the flexibility of traditional receivers.
The resulting architecture is easily scalable to any number of users and is composed of components that are individually interpretable and of reasonable complexity.
We are interested in both uplink and downlink \gls{OFDM} transmissions.
Our solution additionally exploits the \gls{OFDM} structure  to counteract two known drawbacks of conventional receivers that are overlooked in the current literature.
First, it improves the prediction of the channel estimation error statistics, that can only be obtained for the pilot signals in a standard receiver, resulting in largely sub-optimal detection accuracy. 
Motivated by the recent successes of \glspl{CNN} in physical layer tasks~\cite{8752012, honkala2020deeprx}, we use \glspl{CNN} to predict those channel estimation error statistics for every position of the \gls{OFDM} time-frequency grid.
In contrast to the traditional approach that is based on mathematical models, the \glspl{CNN} learn the error statistics from the data during training.
The second improvement is the demapper, which computes \glspl{LLR} over the transmitted bits.
We propose a \gls{CNN}-based demapper that operates on the entire \gls{OFDM} grid, unlike a traditional demapper that operates on individual \glspl{RE}. 
In doing so, our demapper is able to better cope with the residual distortions of the equalized signal.
The resulting \gls{ML}-enhanced receiver is optimized such that all \gls{ML} components are jointly trained to maximize the information rate of the transmission \cite{9118963}.

The proposed architecture is evaluated on \gls{3GPP}-compliant channel models with two different pilot configurations supported by the 5G \gls{NR} specifications. 
Both uplink and downlink transmissions are studied using \gls{TDD}.
We compare the coded \gls{BER} achieved by different schemes for user speeds ranging from \SIrange[range-units = single]{0}{130}{\km\per\hour}.
Two baselines were implemented, the first one being a traditional receiver implementing a \gls{LMMSE} channel estimation and an \gls{LMMSE} equalizer. 
The second baseline also uses \gls{LMMSE} equalization, but is provided with perfect channel knowledge at pilot positions and the exact second order statistics of the channel estimation errors.
Our results show that the gains provided by the \gls{ML}-enhanced receiver increase with the user speed, with small \gls{BER} improvements at low speeds and significant ones at high speeds.
We have also observed that the gains in the uplink are more pronounced than in the downlink, due to channel aging which significantly penalizes the downlink precoding scheme.


\subsection*{Related literature}

Most of the published literature focuses exclusively on the equalization step, which estimates the sent signals from the received ones.
Although performance gains over traditional equalizers were shown, the solutions typically assume perfect \gls{CSI}, are too computationally demanding to be practical, or are trained and evaluated over too simplistic channel models \cite{8642915, 8646357, 9103314, goutay2020deep, pratik2020remimo}.
For example, the OAMP-Net equalizer proposed in \cite{8646357} performs ten matrix inversions to equalize one signal and the MMNet architecture presented in \cite{9103314} needs to be trained for every channel realization. 
Moreover, these papers assume perfect \gls{CSI}, which is not available in practice. For this reason, there is another line of research targeting improved channel estimation \cite{8272484, mashhadi2020pruning, 8752012}.
However, they require ground-truth of the channel realizations during training, which can only be approximated with costly measurement campaigns in practice.
Finally, it has also been proposed to improve the estimation of the bit probabilities by replacing the demapper with a \gls{NN}, but this solution has only been studied for \gls{SISO} setups \cite{shental2020machine, pilotless20, oztekin2019efficient, 9345976}.
In this work, we propose a complete \gls{MU}-\gls{MIMO} receiver that both improves the channel estimation error statistics and the predicted \glspl{LLR} without requiring perfect \gls{CSI} knowledge at training.

More recent papers also leverage a single \gls{NN} to jointly perform multiple processing steps.
This idea has first been proposed in \cite{8052521} to perform joint channel estimation and equalization in a \gls{SISO} setup.
This work has then been extended by additionally learning the demapper and operating directly on time-domain samples~\cite{zhao2018deepwaveform, korpi2020deeprx}.
The DeepRX architecture presented in \cite{honkala2020deeprx} shows impressive results on \gls{SIMO} channels while being 5G compliant.
Another standard-compliant receiver has been proposed for Wi-Fi communications using both synthetic and real-world data \cite{zhang2020deepwiphy}.
Regarding \gls{MIMO} transmissions, a special form of recurrent neural network called reservoir computing has been leveraged in \cite{zhou2019learning} to process time-domain \gls{OFDM} signals.
The DeepRX receiver has also been extended with a so-called transformation layer to handle \gls{MIMO} transmissions \cite{korpi2020deeprx}. 
Their \gls{CNN}-based solution is fed with frequency-domain signals and outputs \glspl{LLR} for the transmitted bits. 
DeepRX MIMO shows important gains on \gls{SU}-\gls{MIMO} channels, but remains very computationally expensive.
Compared to our solution, the main disadvantages of these \gls{NN}-based \gls{MIMO} receivers are their lack of scalability and interpretability. 
They are tailored for a specific number of users or transmit antennas, and the \gls{CSI} needed for downlink precoding can not be easily extracted.

The rest of this paper is structured as follows : In Section~\ref{sec:system_model}, we introduce the channel model and both the uplink and downlink baseline receiver architectures. 
In Section \ref{sec:ml-receiver}, we highlight two limitations of these architectures and detail how we address them using \glspl{CNN}.
Section \ref{sec:evaluations} provides simulation results to compare the various schemes. Section \ref{sec:conclusion} concludes the paper.

\bigskip

\textbf{Notations :} 
$\RR$ and $\CC$ denotes the set of real and complex numbers.
Tensors/matrices are denoted by bold upper-case letters and vectors are denoted by bold lower-case letters.
We denote by $\Tm_{a, b} \in \CC^{N_c \times N_d}$ ($\tv_{a, b, c} \in \CC^{N_d}$, $t_{a, b, c, d} \in \CC$) the matrix (vector, scalar) formed by slicing the tensor $\Tm \in \CC^{N_a \times N_b \times N_c \times N_d}$ along the first two (three, four) dimensions.
The notation $ \Tm^{(k)}$ indicates that the quantity at hand is only considered for the $k^{\text{th}}$ user, and $\vv_{\shortminus a}$ corresponds to the vector $\vv$ from which the $a^{\text{th}}$ element was removed.
$||\Mm||_\text{F}$ denotes the Frobenius norm of $\Mm$.
$\text{Card}(\mathcal{S})$ denotes the number of elements in a set $\mathcal{S}$, $\text{vec}\LB \cdot \RB$ the vectorization operator, and $\odot$ the element-wise product.
$(\cdot)\tp$, $(\cdot)\htp$, and $(\cdot)^*$ denote respectively the transpose, conjugate transpose, and element-wise conjugate operator.
$\Id_N$ is the $N \times N$ identity matrix and $\mathds{1}_{N \times M}$ is the $N\times M$ matrix where all elements are set to $1$.
The imaginary unit  is $j$, such that $j^2 = -1$.

\section{System Model} 
\label{sec:system_model}

We consider a \gls{MU-MIMO} system, where $N_k$ single-antenna users communicate with a \gls{BS} equipped with $N_m$ antennas in the uplink and downlink.
The number of bits transmitted per channel use is denoted by $B$ and $\mathcal{C} = \LP c_1, \cdots, c_{2^{B}} \RP \in \CC^{2^B}$ is the \gls{QAM} constellation used to transmit data.
This section introduces the channel model and the baselines against which the proposed approach is benchmarked.

 \begin{figure*}[t!]
\hspace{20pt}	
  	\begin{subfigure}{0.15\textwidth}
    	\includegraphics[height=125pt]{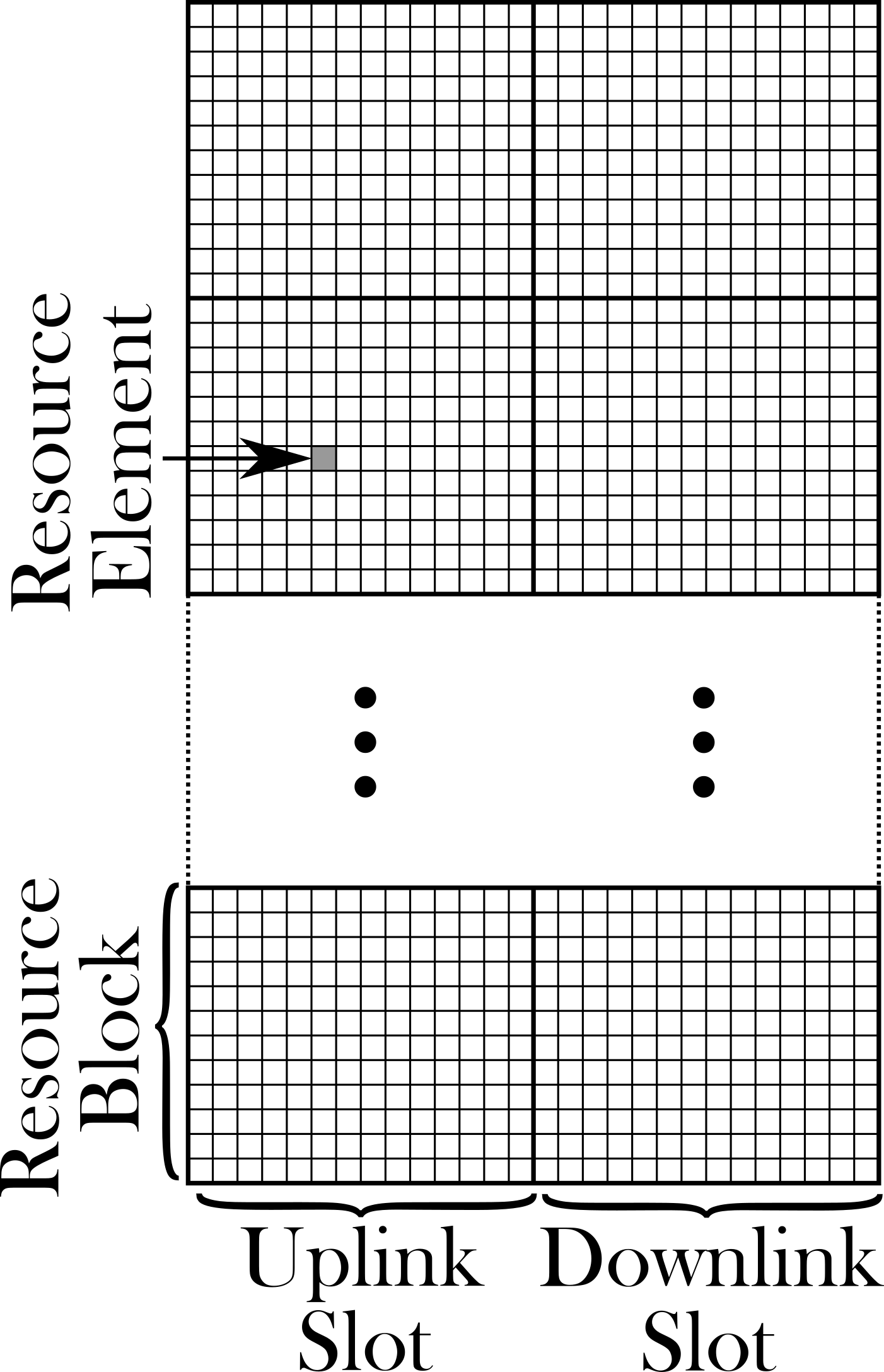} 
  	\caption{Resource grid.}
  	\label{fig:resource_grid}
	\end{subfigure}%
\hspace{35pt}
  	\begin{subfigure}{0.33\textwidth}
  	\includegraphics[height=125pt]{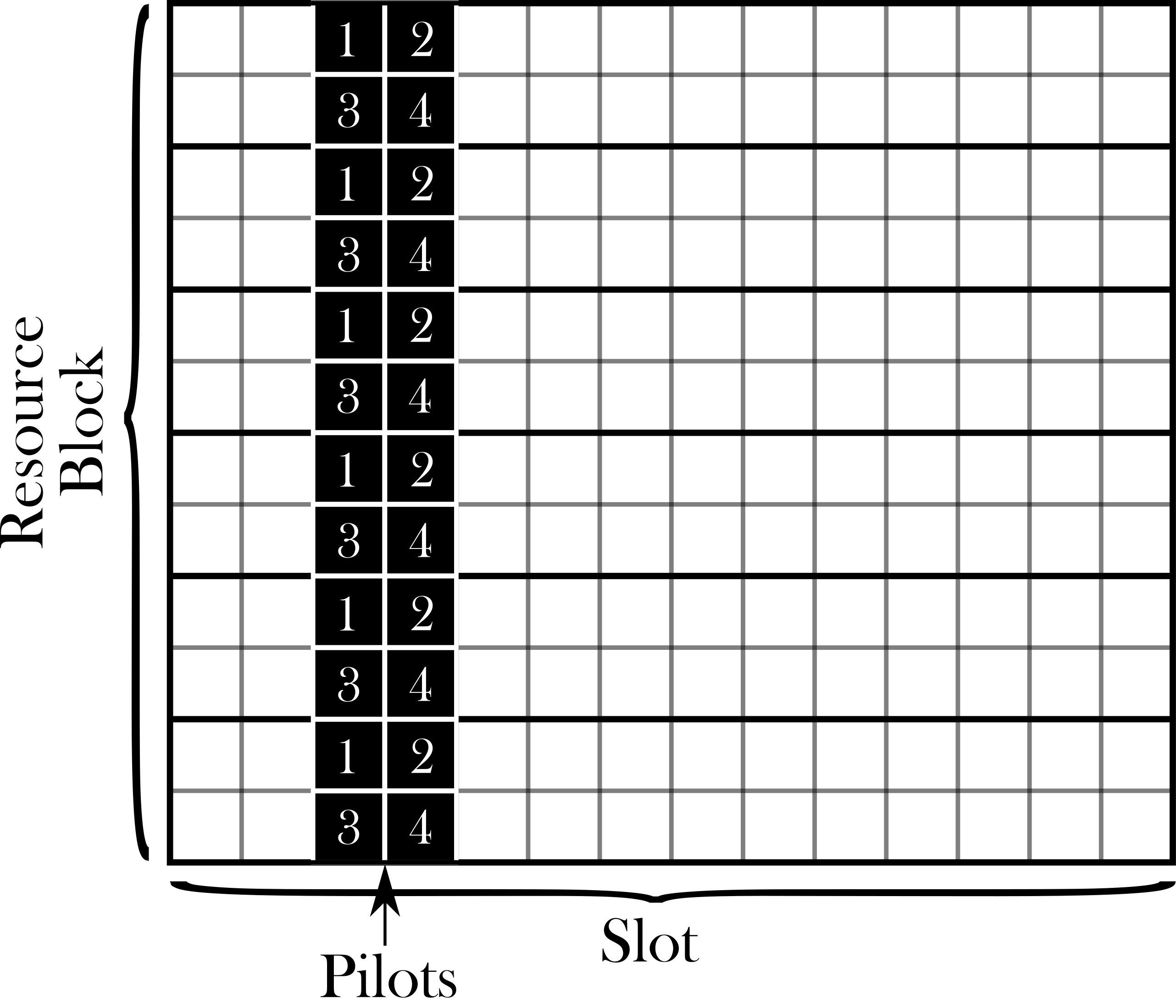}
  	\caption{1P pilot pattern.}
  	\label{fig:1P_pattern}
	\end{subfigure}%
\hspace{5pt}
  	\begin{subfigure}{0.33\textwidth}
  	\includegraphics[height=125pt]{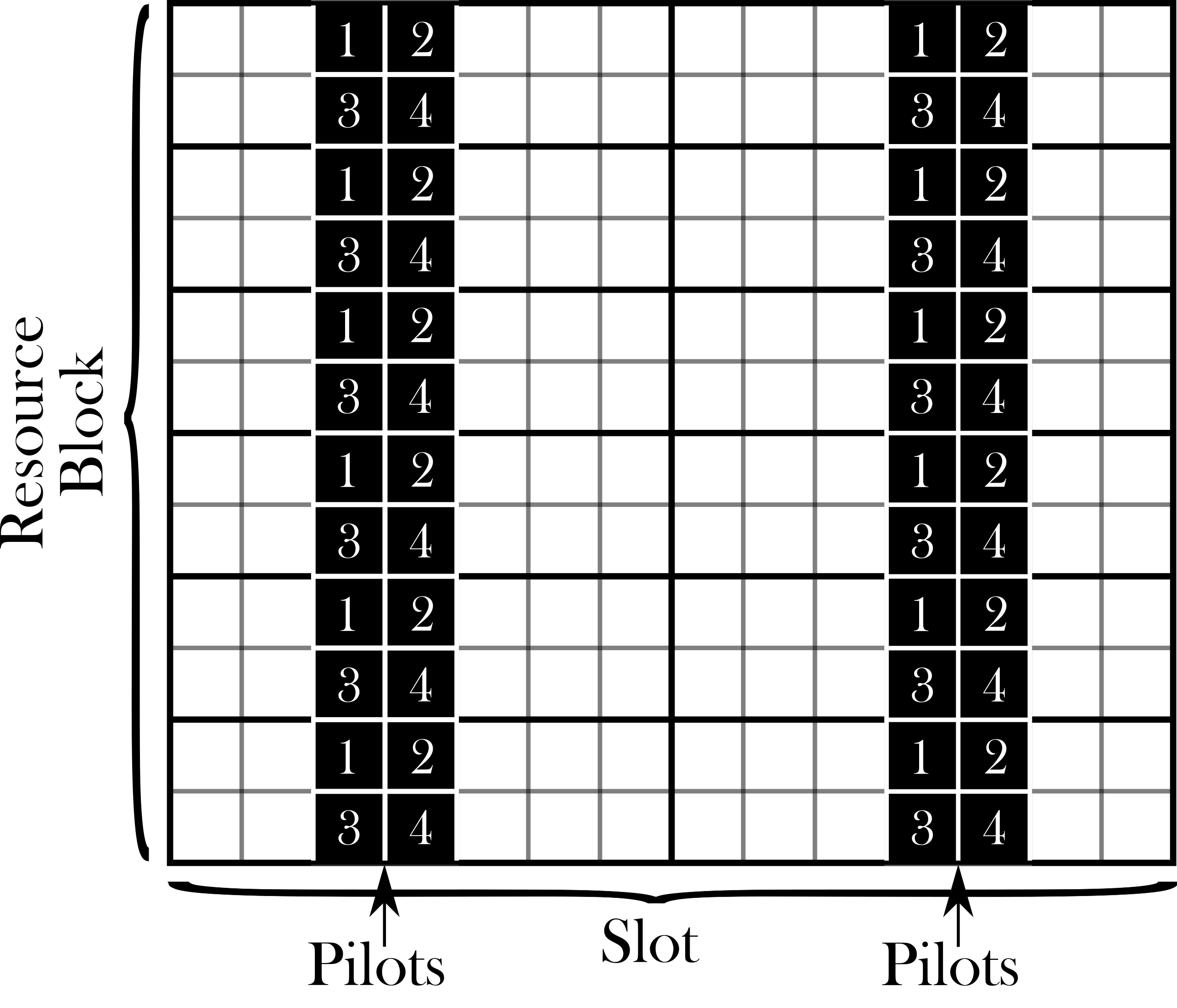}
  	\caption{2P pilot pattern.}
  	\label{fig:2P_pattern}
	\end{subfigure}%
\hspace{20pt}
\caption{Pilots are arranged on the \gls{RG} according to two different patterns, where each number corresponds to a different transmitter.}
\label{fig:channel_model}
\end{figure*}

\subsection{Channel model}

An \gls{OFDM} waveform is considered with $N_f$ subcarriers, divided into resource blocks consisting of  twelve adjacent subcarriers. 
Adjacent \gls{OFDM} symbols are gathered into groups of size $N_t$, referred to as \emph{slots}.
The entire frequency-time grid is called the \gls{RG} and contains all \glspl{RE}, as shown in Fig.~\ref{fig:resource_grid}.
The channel coefficients form a 4-dimensional tensor denoted by $\Hm \in \CC^{N_f \times 2N_t \times N_m \times N_k}$, such that $\Hm_{f, t} \in \CC^{N_m \times N_k}$ is the channel matrix at \gls{RE} $(f,t)$, and $\hv_{f, t, k}\in\CC^{N_m}$ is the channel vector at \gls{RE} $(f,t)$ and for user $k$.
Duplexing is achieved through \gls{TDD}, such that a slot is either assigned to the  uplink or downlink in an alternating fashion, as illustrated in Fig.~\ref{fig:resource_grid}.
More precisely, the first slot is assigned to the uplink and the second slot is assigned to the downlink.
It is assumed that channel reciprocity holds, i.e., $\Hm_{f,t}$ refers as well to the uplink or the downlink channel.
To enable channel estimation, a transmitter sends pilot signals on dedicated \glspl{RE} according to a predefined pilot pattern.
We assume, without loss of generality, that all pilots are equal to one.
Two pilot patterns are considered in this work, referred to as the 1P and 2P pilot patterns, which respectively contain pilots on two or four symbols within a slot.
Fig.~\ref{fig:1P_pattern} and \ref{fig:2P_pattern} respectively show the 1P and 2P pilot patterns over a resource block assuming 4 users.
The set of \glspl{RE} carrying pilots for a user $k \in \LP 1, \dots, N_k \RP$ is denoted by $\Pc^{(k)}$ and the numbers of subcarriers and symbols carrying pilots are respectively denoted by $N_{P_f}$ and $N_{P_t}$.
As an example, if the 1P pattern shown in Fig.~\ref{fig:1P_pattern} is used with $N_f = 12$, the positions $(\text{subcarrier}, \text{symbol})$ of all \glspl{RE} carrying pilots for user 1 are denoted by $\mathcal{P}^{(1)} = \{(1, 3), (3, 3), (5,3), (7, 3), (9, 3), (11,3)\}$, resulting in $N_{P_f} = 6$ and $N_{P_t} = 1$.
Note that when a \gls{RE} is allocated to a user for the transmission of a pilot, other users do not transmit any signal (data nor pilot) on that \gls{RE}.
As a consequence, pilots do not experience any interference.
The noise power is denoted by $\sigma^2$ and assumed equal for all users and all \glspl{RE}.
In the following, perfect power control is assumed over the \gls{RG} such that the mean energy corresponding to a single \gls{BS} antenna and a single user is one, i.e., $\EE\LSB |h_{f,t,m,k}|^2 \RSB = 1$.
The \gls{SNR} of the transmission is defined as 
\begin{equation}
\label{eq:snr}
\text{SNR} = 10 \log{\frac{\EE\LSB |h_{f,t,m,k}|^2 \RSB}{\sigma^2}} = 10 \log{\frac{1}{\sigma^2}} \, [\si{dB}].
\end{equation}

\subsection{Uplink baseline}
\label{sec:baseline_ul}
\begin{figure*}[!t]
    \centering
    \includegraphics[width=0.8\textwidth]{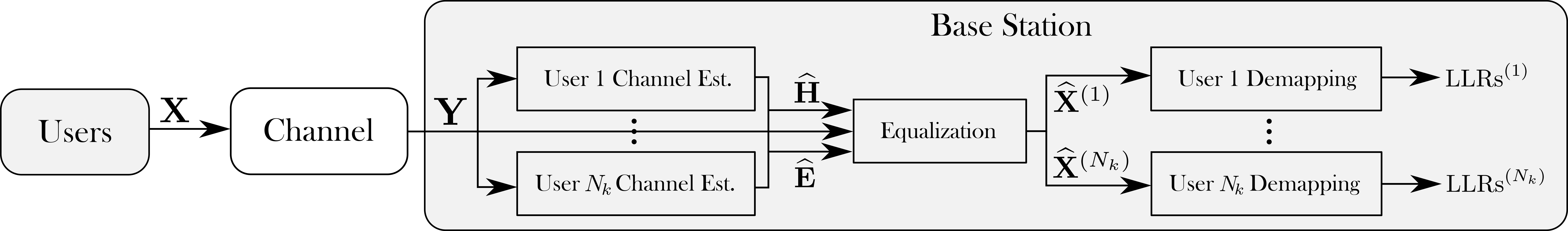}
    \caption{Architecture of the uplink communication system.}
    \label{fig:uplink}
\end{figure*}

In uplink, the \gls{BS} aims to recover the bits transmitted simultaneously by the $N_k$ users on \gls{RE} carrying data. 
The tensors of transmitted and received signals of all users are respectively denoted by $\Xm \in \CC^{N_f \times 2N_t \times N_k}$ and $\Ym \in \CC^{N_f \times 2N_t \times N_m}$. 
In this scenario, only the uplink slot is used and therefore all signals with indices $t > N_t$ are ignored and the corresponding values are set to $0$.
The transfer function of the uplink channel for a \gls{RE} $(f,t)$ is
\begin{equation}
\label{eq:transfert_function}
\yv_{f,t} = \Hm_{f,t} \xv_{f,t} + \nv_{f,t}
\end{equation}
where $\nv_{f,t} \sim \Cc\Nc \LB \zerov, \sigma^2\Id_{N_m} \RB$ is the noise vector.
The architecture of the uplink system is shown in Fig.~\ref{fig:uplink}, where the IFFT (FFT) operation and the addition (removal) of the cyclic prefix before (after) the channel are not shown for clarity.
The channel estimation, equalization, and demapping stages of the baseline will be explained in the following sections.

\subsubsection{Channel estimation}
\label{sec:ch_est_1}

As the pilots are assumed to be orthogonal, \gls{LMMSE} channel estimation can be carried out for each user independently.
The channel covariance matrix providing the spatial, temporal, and spectral correlations between all \glspl{RE} carrying pilots is denoted by $\Sigmam \in \CC^{N_{P_f}N_{P_t}N_m \times N_{P_f}N_{P_t}N_m}$.
In the following, it is assumed that the precise local statistics of the receivers are not available.
The channel and receiver statistics are therefore averaged over the entire cell, resulting in a zero-mean channel, and a discussion on how these statistics can be obtained is provided in Section~\ref{sec:stats}.
For a user $k \in \{1,\dots,N_k\}$, the \gls{LMMSE} channel estimate at \glspl{RE} carrying pilots is denoted by $\widehat{\Hm}^{(k)}_{\mathcal{P}^{(k)}} \in \CC^{N_{P_f} \times N_{P_t} \times N_m}$ and given by
\begin{equation}
\displaystyle
\text{vec}\LB\widehat{\Hm}^{(k)}_{\mathcal{P}^{(k)}}\RB = \Sigmam \LB \Sigmam  + \sigma^2\Id_{N_{P_f}N_{P_t}N_m } \RB^{-1} \text{vec}\LB\Ym^{(k)}_{\mathcal{P}^{(k)}}\RB
\end{equation}
where $\Ym^{(k)}_{\mathcal{P}^{(k)}} \in \CC^{N_{P_f} \times N_{P_t} \times N_m}$ is the tensor of received pilots for user $k$.
Channel estimation could also be performed at \glspl{RE} carrying data \cite{pilotless20}, but this would require knowledge of the channel statistics at those \glspl{RE}, which are typically not available in practice.

Inspired by the \gls{3GPP} guidelines \cite{3gpp.36.141}, the channel estimates for all \glspl{RE} are computed by first linearly interpolating the estimates from \glspl{RE} carrying pilots in the frequency dimension and then using the estimate at the \gls{NIRE} on the neighboring \glspl{RE}.
It is also possible to leverage temporal linear interpolation between the \gls{OFDM} symbols carrying pilots when the 2P pilot pattern is used.
The so-obtained tensor of channel estimates is denoted by $\widehat{\Hm}^{(k)} \in \CC^{N_f \times 2N_t \times N_m}$. 
The overall channel estimation for all users $\widehat{\Hm} \in \CC^{N_f \times 2N_t \times N_m \times N_k} $ is obtained by stacking the channel estimates of all users.
Since only the uplink slot is considered here, the channel estimates for the last $N_t$ symbols (downlink slot) are set to be null.
The channel estimation error is denoted by $\widetilde{\Hm}$ and is such that $\Hm = \widehat{\Hm} + \widetilde{\Hm}$.
For a \gls{RE} $(f,t)$, we define
\begin{equation}
\label{eq:E}
\Em_{f,t} \coloneqq \EE \LSB \widetilde{\Hm}_{f, t} \widetilde{\Hm}_{f, t}\htp \RSB
\end{equation}
as the sum of the \emph{spatial} channel estimation error covariance matrices from all users.

\begin{figure*}[b!]
\begin{minipage}[b]{0.65\textwidth}
\includegraphics[height=105pt,left]{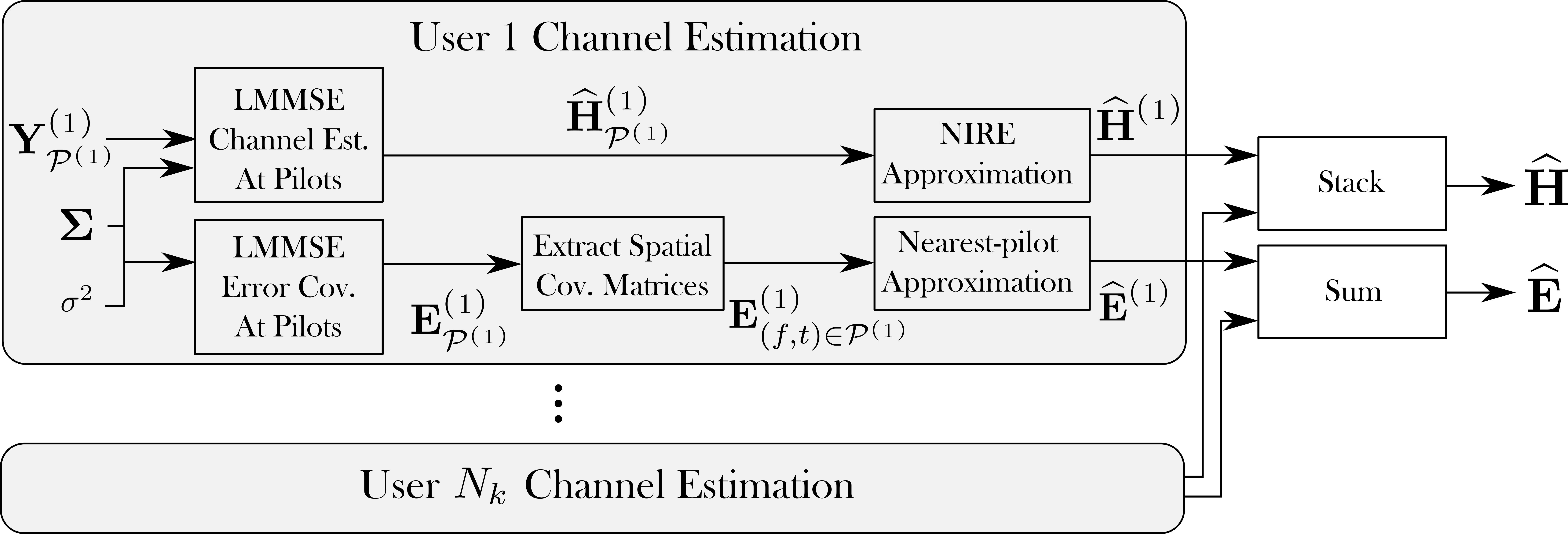}
  \caption{Uplink channel estimation.}
    \label{fig:ch_est_tradi}
\end{minipage}
\begin{minipage}[b]{0.35\textwidth}
\includegraphics[height=78pt,right]{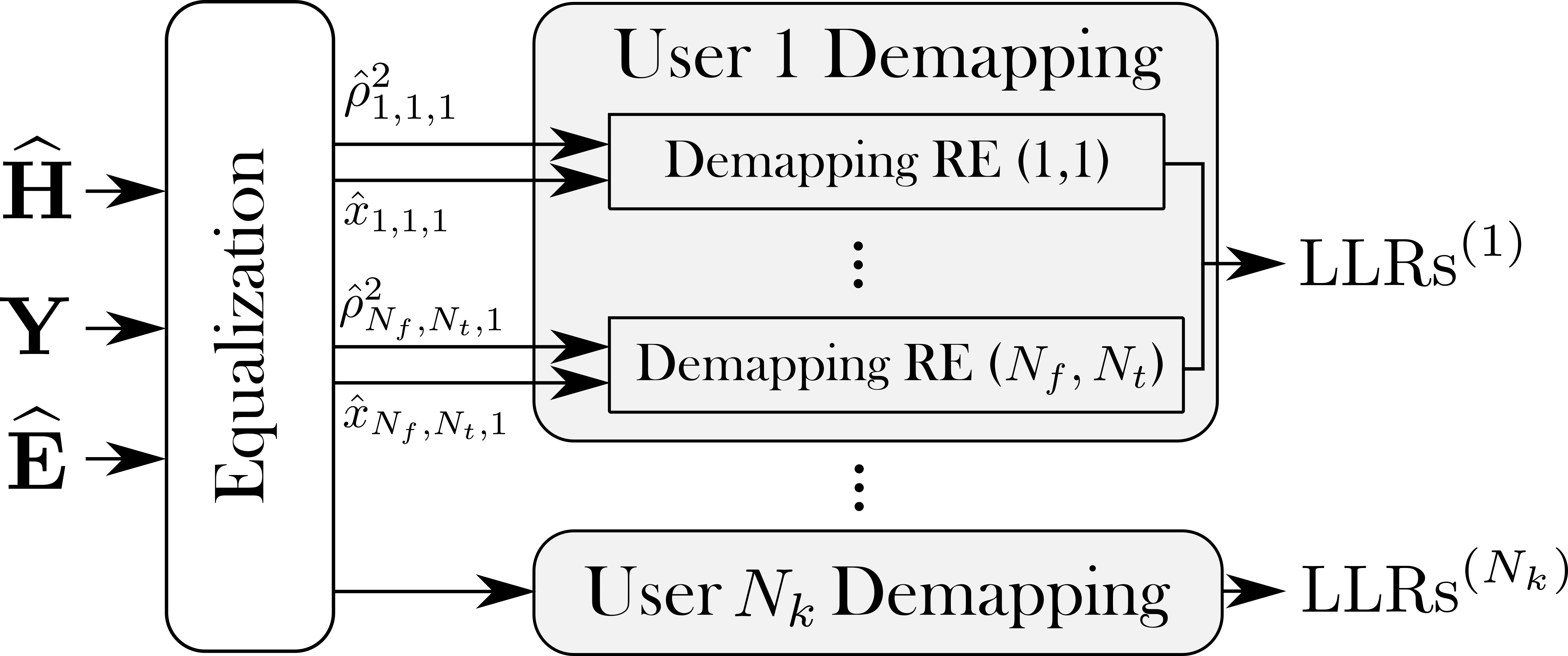}
\vspace{8pt}
  \captionof{figure}{Uplink demapping}
  \label{fig:eq_dmp_tradi}
  \end{minipage}
\end{figure*}

\subsubsection{Equalization}

The widely used \gls{LMMSE} equalizer is leveraged, as it maximizes the post-equalization \gls{SNR} among all linear operators.
However, as computing a dedicated \gls{LMMSE} operator for each \gls{RE} is infeasible in practice due to prohibitive complexity, we resort to a grouped-\gls{LMMSE} equalizer, i.e., a single \gls{LMMSE} operator is applied to a group of adjacent \glspl{RE} spanning multiple subcarriers $f \in \{F_b,\dots, F_e\}$ and symbols $t \in \{T_b,\dots,T_e\}$.
Under this assumption, the \gls{LMMSE} operator for the group of \glspl{RE} spanning $\{F_b,\dots, F_e\}\times\{T_b,\dots,T_e\}$ is 

\begin{align}
\displaystyle
\label{eq:W}
\Wm_{f,t} = & \LB \sum_{f'=F_b}^{F_e} \sum_{t'=T_b}^{T_e}  \widehat{\Hm}_{f', t'}\htp \RB \\
&\LB \sum_{f'=F_b}^{F_e} \sum_{t'=T_b}^{T_e}  \widehat{\Hm}_{f', t'} \widehat{\Hm}_{f',t'}\htp + \Em_{f',t'} + \sigma^2 \Id_{N_m}  \RB^{-1} \nonumber .
\end{align}
where $\Wm_{f,t} \in \CC^{N_k \times N_m}$ is constant over $\{F_b,\dots, F_e\}\times\{T_b,\dots,T_e\}$ (as derived in Appendix~\ref{app:LMMSE}).

 The post-equalization channel is expected to be an additive noise channel. More precisely, for any \gls{RE} $(f,t)$, the demapper expects the output of the equalizer $\hat{\xv}_{f, t} \in \CC^{N_k}$ to be such that $\hat{\xv}_{f,t} = \xv_{f,t} + \zv_{f,t}$ where $\zv_{f,t}$ is an additive noise term.
However, this decomposition is not achieved by the \gls{LMMSE} equalizer (see Section 1.6.1 of \cite{heath2018foundations} for a more detailed discussion).
To obtain such a post-equalized channel, the following diagonal matrix is applied to the output of the \gls{LMMSE} equalizer
\begin{equation}
\Dm_{f,t} = \LB  \LB \Wm_{f,t} \hat{\Hm}_{f, t} \RB \odot \Id_{N_k} \RB^{-1}
\end{equation}
which re-scales the equalizer output so that the post-equalization \gls{SNR} remains maximized.
For a \gls{RE} $(f,t) \in \{F_b,\dots, F_e\}\times\{T_b,\dots,T_e\}$, the equalized vector $\hat{\xv}_{f, t} \in \CC^{N_k}$ is computed by
\begin{equation}
\hat{\xv}_{f, t} = \Dm_{f,t} \Wm_{f,t} \yv_{f,t}.
\end{equation}
The equalized symbols of user $k$ are denoted by $\widehat{\Xm}^{(k)} \in \mathbb{C}^{N_f \times N_t}$, as shown in Fig.~\ref{fig:uplink}.

\subsubsection{Demapping} 
\label{sec:demapping_ul}

For a \gls{RE} $(f,t)$, let us denote by $\wv_{f,t,k}$ the column vector made of the $k^{\text{th}}$ line of the matrix $\Wm_{f,t}$ and by $\Hm_{f, t, -k}$ the tensor made of the channel coefficients of all users except user $k$. 
After equalization, the uplink channel can be viewed as $N_f N_t N_k$ parallel additive noise channels that can be demodulated independently for every \gls{RE} and every user.
For a \gls{RE} $(f,t)$ and user $k$, the post-equalization channel is expressed as

{\small
\begin{equation}
\hat{x}_{f, t, k}= x_{f, t, k} + \underbrace{\frac{\wv_{f, t, k}\tp \LB \widehat{\Hm}_{f, t, \shortminus k} \xv_{f, t, \shortminus k} + \widetilde{\Hm}_{f, t} \xv_{f, t} +  \nv_{f, t} \RB}{\wv_{f, t, k}\tp \widehat{\hv}_{f, t, k} }}_{z_{f, t, k}}
\end{equation}
}
where the noise $z_{f, t, k}$ includes both the interference and the noise experienced by user $k$.
Its variance is given by
\begin{align}
\displaystyle
\label{eq:eq_snr_ul}
\rho_{f, t, k}^2 & =  \EE\LSB z_{f, t, k}^* z_{f, t, k} \RSB\\
& = \frac{\wv_{f, t, k}\htp \LB\widehat{\Hm}_{f, t, \shortminus k} \widehat{\Hm}_{f, t, \shortminus k}\htp + \Em_{f,t} + \sigma^2 \Id_{N_m} \RB \wv_{f, t, k}}{\wv_{f, t, k}\htp \widehat{\hv}_{f, t, k} \widehat{\hv}_{f, t, k}\htp \wv_{f, t, k}} \nonumber .
\end{align}
We denote by $\mathcal{C}_{b,0}$~($\mathcal{C}_{b,1}$) the subset of $\mathcal{C}$ which contains all symbols with the $b^{\text{th}}$ bit set to 0~(1).
Assuming the noise $z_{f,t,k}$ is Gaussian\footnote{This is not true in general as the interference and channel estimation errors are not Gaussian distributed.}, the \glspl{LLR} of the $b^{\text{th}}$ bit transmitted by user $k$ on the \gls{RE} $(f, t)$ is given by

\vspace{-\baselineskip}
{\small 
\begin{equation}
\text{LLR}_{f, t, k}^{\text{UL}}(b) = \ln{\frac{
\sum_{c\in\mathcal{C}_{b,1}} \exp{  - \frac{1}{\rho_{f, t, k}^2} \abs{ \hat{x}_{f, t, k} - c}^2 }
}{
\sum_{c\in\mathcal{C}_{b,0}} \exp{  - \frac{1}{\rho_{f, t, k}^2} \abs{\hat{x}_{f, t, k} - c}^2 }
} } .
\end{equation}
}
The equalization and demapping process is schematically shown in Fig.~\ref{fig:eq_dmp_tradi}.

\subsection{Downlink baseline}
\label{sec:baseline_dl}

\begin{figure*}[t]
    \centering
    \includegraphics[width=0.8\textwidth]{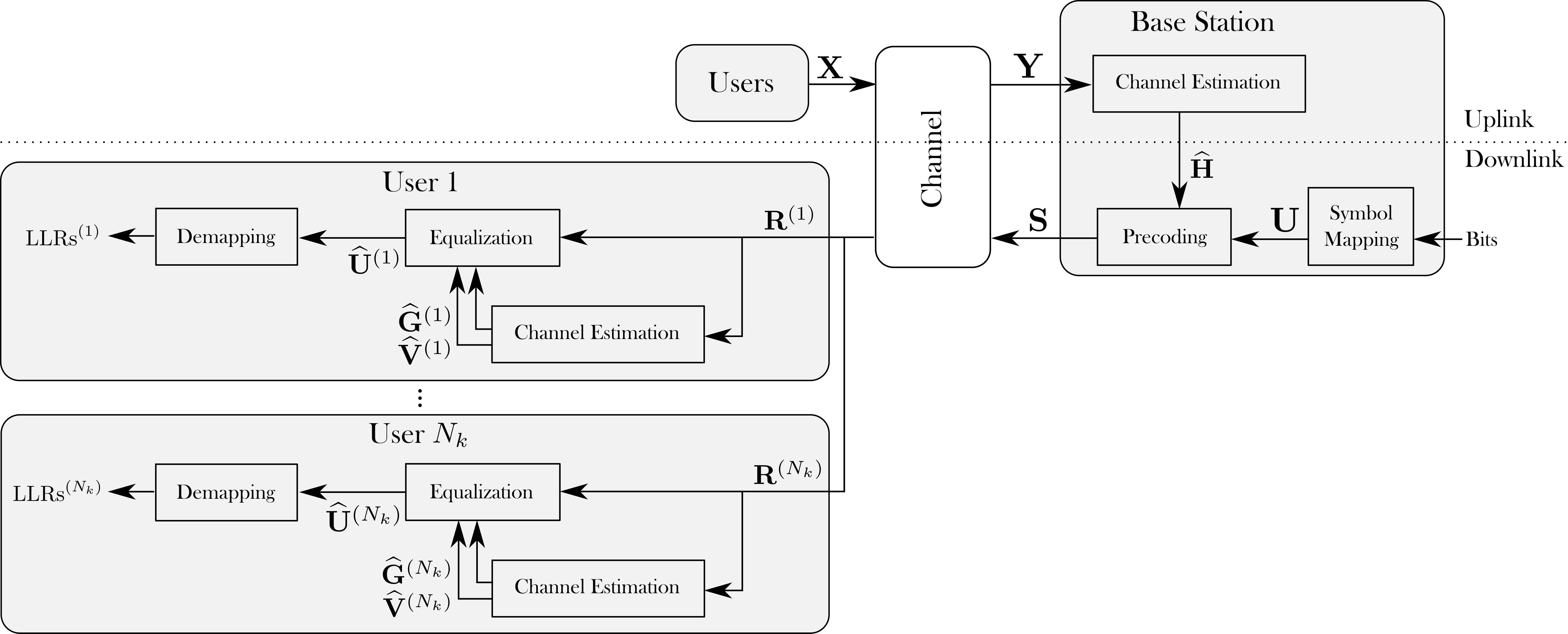}
    \caption{Architecture of the downlink communication system.}
    \label{fig:downlink}
\end{figure*}

The \gls{BS} aims to simultaneously transmit to  $N_k$ users on all \glspl{RE} of the downlink slot. 
The signal transmitted by the \gls{BS} is precoded to mitigate interference.
We remind that downlink transmissions occur after the uplink slot, as shown in Fig.~\ref{fig:resource_grid}.
Let us denote by $\Um \in \mathcal{C}^{N_f \times 2N_t \times N_k}$ and by $\Sm \in \CC^{N_f \times 2N_t \times N_m}$ the tensors of unprecoded and precoded symbols, respectively.
We denote by $\Rm \in \CC^{N_f \times 2 N_t \times N_k}$ the tensor of symbols received by the $N_k$ users.
Those quantities are only relevant on the downlink slot and therefore are considered null on the first $N_t$ symbols, i.e., $\uv_{f,t} = \sv_{f,t} = \rv_{f,t} = \mathbf{0} \quad  \forall (f,t) \in \{1, \dots, N_f\} \times \{1, \dots, N_t\}$.
The downlink transfer function of the channel  for a \gls{RE} $(f,t)$ is
\begin{equation}
\label{eq:down_chan}
\rv_{f,t} = \Hm_{f,t}\htp \sv_{f,t} + \qv_{f,t}
\end{equation}
where $\qv_{f,t} \sim \Cc\Nc(\zerov,\sigma^2 \Id_{N_k})$ is the noise vector,  considered null in the first $N_t$ symbols.
For convenience, the noise variance $\sigma^2$ is assumed to be the same as in the uplink.
Fig.~\ref{fig:downlink} shows the architecture of the downlink system, where the IFFT (FFT) operation and the addition (removal) of the cyclic prefix before (after) the channel are again not shown for clarity.
In the rest of this section, we detail the precoding, channel estimation and equalization, and demapping steps.

\subsubsection{Precoding}

Precoding requires estimation of the downlink channel. 
As \gls{TDD} is used, we can exploit channel reciprocity so that the downlink channel can be estimated using the nearest-pilot approach, i.e., $\widehat{\Hm}_{f, N_t + 1 \leq t \leq 2N_t} = \widehat{\Hm}_{f,N_t}$.
Precoding is achieved by exploiting the uplink-downlink duality~\cite{viswanath2003sum}, which results in using $\Wm_{f,t}\htp$ as precoding matrix that can be computed using~\eqref{eq:W}.
Normalization is performed to ensure that the average energy per transmitted symbol equals one, by applying the diagonal matrix 
\begin{equation}
\Nm_{f, t} = \LB \LB \Wm_{f,t}\Wm_{f,t}\htp \RB \odot \Id_{N_k} \RB^{-\frac{1}{2}}
\end{equation}
leading to the precoded signal
\begin{equation}
\sv_{f,t}  = \Nm_{f,t} \Wm_{f, t}\htp  \uv_{f,t}.
\end{equation}
The channel transfer function~\eqref{eq:down_chan} can be rewritten as
\begin{equation}
\rv_{f,t} = \underbrace{\Hm_{f,t}\htp \Nm_{f,t} \Wm_{f, t}\htp }_{\Gm_{f,t}} \uv_{f,t} + \qv_{f,t} 
\end{equation}
where $\Gm_{f, t}  \in \CC^{N_k \times N_k}$ is referred to as the \emph{equivalent} downlink channel for the \gls{RE} $(f,t)$.
Each user $k$ receives its signal $r^{(k)}_{f,t}$  and the corresponding channel, i.e., the $k^{\text{th}}$ row of $\Gm_{f, t}$, is denoted by ${\gv^{(k)}_{f,t}}\tp \in \mathbb{C}^{N_k}$.
Finally, the equivalent channel experienced by user $k$ for the entire \gls{RG} is denoted  by $\Gm^{(k)} \in \mathbb{C}^{N_f \times 2N_t \times N_k}$.

\subsubsection{Channel estimation and equalization}
\label{sec:dl_ch_est}

To enable estimation of the equivalent  downlink channel by the users, pilot signals are transmitted by the \gls{BS} using the same pilot patterns as in the uplink (Fig.~\ref{fig:channel_model}).
Each user $k$ estimates its equivalent channel $\widehat{\Gm}^{(k)} \in \mathbb{C}^{N_f \times 2N_t \times N_k}$, where for a given \gls{RE} $(f,t)$, the element $\hat{g}^{(k)}_{f,t,k}$ corresponds to the main channel coefficient, whereas the elements $\hat{g}^{(k)}_{f,t,i}, i \neq k$ correspond to the interference channel coefficients.
As in the uplink, \gls{LMMSE} estimation, followed by spectral and possibly temporal interpolation, is used, but it is assumed that the elements of ${\hat{\gv}^{(k)}_{f,t}}$ are uncorrelated.
Therefore, channel estimation is performed independently for the main channel and each interference channel, enabling easy scalability to any number of interferers.
The covariance matrices used to estimate the main channel and one of the interfering channels of a given user are denoted by $\Omegam \in \CC^{N_{P_f}N_{P_t} \times N_{P_f}N_{P_t}} $ and by $\Psim \in \CC^{N_{P_f}N_{P_t} \times N_{P_f}N_{P_t}}$, respectively, and are equal for all users and interferering channels.
The tensor of the equivalent channel estimation error for user $k$ is denoted by $\widetilde{\Gm}^{(k)} \in \CC^{N_f \times 2N_t \times N_k}$, and is such that $\Gm^{(k)} = \widehat{\Gm}^{(k)} + \widetilde{\Gm}^{(k)}$.
The estimation error variances for the main and the $i^{\text{th}}$ interfering channel of user $k$ are respectively denoted by $v^{(k)}_{f,t,k} \coloneqq \EE \LSB \abs{\tilde{g}^{(k)}_{f,t,k}}^2 \RSB$ and $v^{(k)}_{f,t,i} \coloneqq \EE \LSB \abs{\tilde{g}^{(k)}_{f,t,i}}^2 \RSB$.
Similarly, we denote by $\Vm^{(k)} \in\mathbb{R}^{N_f, \times N_t \times N_k}$ the tensor of estimated error variances for user $k$, as shown in Fig.~\ref{fig:downlink}.
An estimation of the transmitted unprecoded symbol for user $k$ is computed by equalizing the received signal as follows

\begin{equation}
\hat{u}^{(k)}_{f, t} = \frac{r^{(k)}_{f, t}}{\hat{g}^{(k)}_{f,t,k}}.
\end{equation}

\subsubsection{Demapping}

The post-equalization downlink channel can be seen as $N_f \times N_t \times N_k$ parallel additive noise channels.
More precisely, for a user $k \in \{1,\dots,N_k\}$ and \gls{RE} $(f,t)$,
\begin{equation}
\label{eq:eq_eq_ch_dl}
\hat{u}^{(k)}_{f, t}
= u_{f, t, k}
+ \underbrace{\frac{\tilde{g}^{(k)}_{f, t, k} u_{f, t, k} 
+ {\gv^{(k)}_{f,t, \shortminus k}}\tp ~ \uv_{f, t, \shortminus k}
+ q_{f, t, k}}{\hat{g}^{(k)}_{f, t, k}}}_{o_{f, t, k}}
\end{equation}
where $o_{f,t,k}$ comprises the channel noise and interference and has variance
\begin{align}
\displaystyle
\label{eq:eq_snr_dl}
\tau_{f, t, k}^2
& =  \EE\LSB o_{f, t, k}^* o_{f, t, k} \RSB  \\
& = \frac{v^{(k)}_{f,t,k} + \hat{\gv}^{(k)~~^{\scriptstyle \mathsf{H}}}_{f, t,\shortminus k} ~ \hat{\gv}^{(k)}_{f,t,\shortminus k}
+ \sum_{i=1, i\neq k}^{N_k} v^{(k)}_{f,t,i} + \sigma^2 }{\abs{\hat{g}^{(k)}_{f, t, k}}^2} \nonumber .
\end{align}
Assuming $o_{f,t,k}$ is Gaussian distributed\footnote{Similarly to the uplink scenario, this is not true in general.}, the \gls{LLR} for the $b^{\text{th}}$ bit transmitted to user $k$ on \gls{RE} $(f, t)$ is given by

\vspace{-\baselineskip}
{\small
\begin{equation}
\displaystyle
\text{LLR}_{f, t, k}^{\text{DL}}(b) = \ln{\frac{
\sum_{c\in\mathcal{C}_{b,1}} \exp{  - \frac{1}{\tau_{f, t, k}^2} \abs{\hat{u}_{f, t, k} - c}^2 }
}{
\sum_{c\in\mathcal{C}_{b,0}} \exp{  - \frac{1}{\tau_{f, t, k}^2} \abs{\hat{u}_{f, t, k} - c}^2 }
} } .
\end{equation}}

\subsection{Estimation of the required statistics}
\label{sec:stats}

The baselines described above require the knowledge of the covariance matrices $\Sigmam$, $\Omegam$, and $\Psim$ which provide the spatial, time, and spectral correlations between the \glspl{RE} carrying pilots.
These matrices can be set based on models or can be empirically estimated by constructing large datasets of uplink, downlink, and interfering pilot signals, as is done in this paper.
The channel estimation error covariances $\Em_{f,t}$, defined in~\eqref{eq:E}, also need to be estimated to compute both the uplink equalization matrices $\Wm_{f,t}$ and the downlink precoding matrices $\Wm_{f,t}\htp$.
Focusing on \glspl{RE} carrying pilots, the estimation error covariance for a user $k$ is given by
\begin{align}
\label{eq:cov_err}
\begin{split}
\Em^{(k)}_{\mathcal{P}^{(k)}}
& =   \EE\LSB \text{vec}\LB \widetilde{\Hm}^{(k)}_{\mathcal{P}^{(k)}}\RB  \text{vec}\LB \widetilde{\Hm}^{(k)}_{\mathcal{P}^{(k)}}\RB\htp \RSB \\
& = \Sigmam - \Sigmam \LB \Sigmam  + \sigma^2 \RB^{-1}\Sigmam 
\end{split}
\end{align}
where $\widetilde{\Hm}^{(k)}_{\mathcal{P}^{(k)}}$ is the channel estimation error at \glspl{RE} carrying pilots.
However, we are only interested in the \emph{spatial} channel estimation error correlations, whereas~\eqref{eq:cov_err} provides the correlations of channel estimation errors between all the receive antennas, subcarriers, and symbols.
For a single pilot position $(f, t) \in \mathcal{P}^{(k)}$, this spatial correlation matrix is defined by
\begin{equation}
\displaystyle
\Em^{(k)}_{(f,t) \in \mathcal{P}^{(k)}} = \EE \LSB  \LB \widetilde{\hv}^{(k)}_{(f,t) \in \mathcal{P}^{(k)}} \RB \LB \widetilde{\hv}_{(f,t) \in \mathcal{P}^{(k)}}^{(k)}\RB ^\mathsf{H} \RSB \in  \CC^{N_m \times N_m}
\end{equation}
and can be extracted from $\Em^{(k)}_{\mathcal{P}^{(k)}}$.
To estimate $\Em_{f,t}$ for \glspl{RE} carrying data, a nearest-pilot approach is leveraged, which sets the value $\Em_{f,t}$ for a \gls{RE} carrying a data signal to the one of the nearest \gls{RE} carrying a pilot signal.
The so-obtained estimation is denoted by $\widehat{\Em}_{f,t}^{(k)}$ for a \gls{RE} $(f,t)$.
The overall spatial estimation error covariance matrix for any \gls{RE} $(f,t)$ is obtained by summing the estimations for all users:
\begin{equation}
\widehat{\Em}_{f,t}
= \sum_{u=1}^{N_k} \widehat{\Em}^{(k)}_{f,t} \quad \in \CC^{N_m \times N_m}.
\end{equation}
The uplink channel and error covariance estimations are depicted in Fig.~\ref{fig:ch_est_tradi}.

In the downlink, the estimation error variances $v^{(k)}_{f,t,k}$ and $v^{(k)}_{f,t,i}, i \neq k$ for user $k$ are estimated following a similar procedure, but with only one receive antenna and using the downlink covariances matrices $\Omegam$ and $\Psim$. 
The resulting quantity is denoted by $\widehat{\Vm}^{(k)}$, as shown in Fig.~\ref{fig:downlink}.
\section{ML-Enhanced receiver architecture}
\label{sec:ml-receiver}

The baselines presented in the previous section have several limitations. 
Especially, the \gls{NIRE} approximation leads to high channel estimation errors for \glspl{RE} that are far from pilots.
Similarly, the grouped-equalization can be inaccurate at those \glspl{RE}.
This section details the architecture of a receiver that builds on the presented baseline and uses  multiple \glspl{CNN} to improve its performance.

\subsection{Receiver training}

\begin{figure*}[t!]
    \centering
    \includegraphics[width=0.8\textwidth]{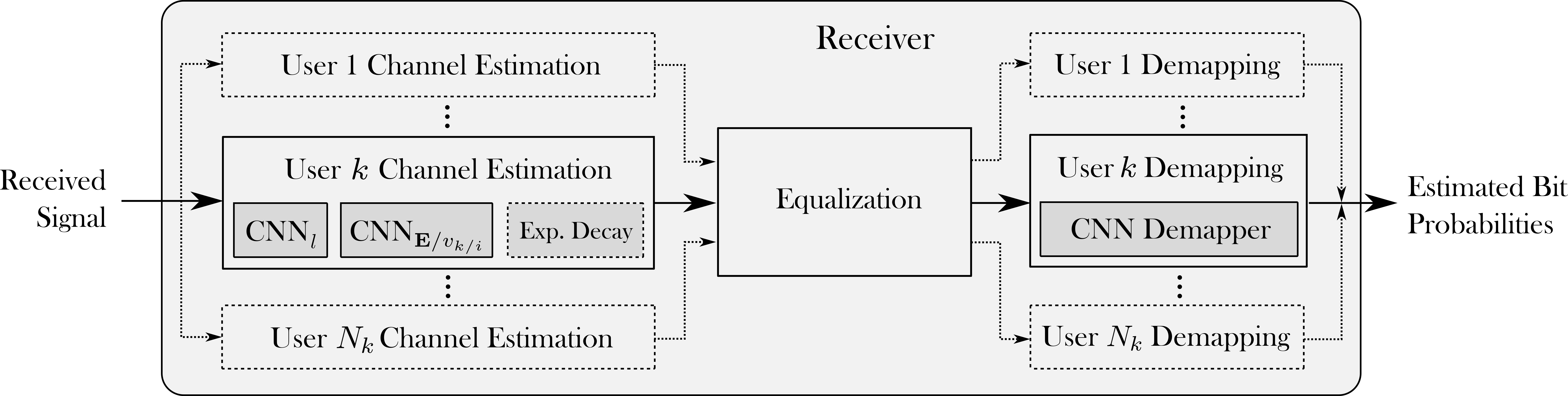}
    \caption{ML-enhanced receiver architecture. The dotted elements are only present in the uplink, where the \gls{BS} jointly processes all users. The dark grey elements are trainable components.}
    \label{fig:ml_receiver}
\end{figure*}

The \gls{ML}-enhanced receiver architecture is shown in Fig.~\ref{fig:ml_receiver}, where the trainable components are represented in dark grey.
In the downlink, each user $k$ only performs the channel estimation, equalization, and demapping of its own signal, and the corresponding components are illustrated with continuous outlines.
However, in the uplink, the \gls{BS} processes all users in parallel, and the additional components are delimited with dotted lines.
Although the internal processing of the trainable components might not be interpretable, using multiple \gls{ML}-based blocks to perform precise and relatively simple signal processing tasks allows to precisely control which parts of the receiver are enhanced.
Moreover, this approach makes the output of each \gls{ML} components easier to interpret, as discussed in Section \ref{sec:E_comp_speed} for the error statistics $\widehat{\Em}$.

Scalability is achieved by using different copies of the same \gls{ML} components for every users, where all copies share the same set of trainable parameters.
We propose to jointly optimize all these components based only on the estimated bit probabilities, and not by training each of them individually. 
This approach is practical as it does not assume knowledge of the channel coefficients at training, that can only be estimated through extensive measurement campaigns for practical channels.
Let's denote by $\thetav$ the set of trainable parameters of the \gls{ML}-enhanced receiver, and by  $\mathscr{b}_{f,t,k,b}$ the sent bit $(f,t,k,b)$.
In the uplink, those parameters are optimized to minimize the total \gls{BCE} :

\vspace{-\baselineskip}
{\small
\begin{align}
\displaystyle
\label{eq:bce}
\mathcal{L} \triangleq & - \sum_{k=1}^{N_k} \sum_{(f, t)\in \mathcal{D}} \sum_{b=1}^{B} \EE_{\mathscr{b}, \Ym} \left[ \mathscr{b}_{f,t,k,b}  \text{log}_2 \LB 
\widetilde{P}_{\thetav} \LB \mathscr{b}_{f,t,k,b} =1 | \Ym \RB \RB \right. \nonumber \\
 &+ \left. \LB 1 - \mathscr{b}_{f,t,k,b}\RB  \text{log}_2 \LB 1 - 
\widetilde{P}_{\thetav} \LB \mathscr{b}_{f,t,k,b} =1 | \Ym \RB \RB
 \right]
\end{align}}
\hspace{-3pt}where $\mathcal{D}$ denotes the set of \glspl{RE} carrying data and $\widetilde{P}_{\thetav} \LB \cdot | \Ym \RB $ is the receiver estimate of the posterior
distribution on the bits given $\Ym$. 
The estimated posterior probabilities are obtained by applying the sigmoid function to the corresponding \glspl{LLR}, i.e., $\widetilde{P}_{\thetav} \LB \mathscr{b}_{f,t,k,b} =1 | \Ym \RB = \text{sigmoid}\LB\text{LLR}_{f,t,k}(b)\RB $.
In the downlink, the receiver parameters are optimized in a similar manner, except that the signal received by the users is $\Rm$.
The expectation in \eqref{eq:bce} is estimated through Monte Carlo sampling using batches of $S$ samples :

\vspace{-\baselineskip}
{\small
\begin{align}
\label{eq:loss_mc}
\displaystyle
\mathcal{L} \approx & - \frac{1}{S} \sum_{s=1}^{S} \sum_{k=1}^{N_k} \sum_{(f, t)\in \mathcal{D}} \sum_{b=1}^{B} \LB \mathscr{b}_{f,t,k,b} \text{log}_2 \LB 
\widetilde{P}_{\thetav} \LB \mathscr{b}_{f,t,k,b}^{[s]} = 1| \Ym^{[s]} \RB \RB \right. \nonumber \\
& \left. + \LB 1 -  \mathscr{b}_{f,t,k,b} \RB \text{log}_2 \LB  1 -
\widetilde{P}_{\thetav} \LB \mathscr{b}_{f,t,k,b}^{[s]} = 1| \Ym^{[s]} \RB \RB \RB
\end{align}}
\hspace{-3pt}where the superscript $[s]$ refers to the $s^{\text{th}}$ sample in the batch.
The loss \eqref{eq:bce} can be redefined as
\begin{equation}
\mathcal{L} = \sum_{k=1}^{N_k} \LB \text{Card}(\mathcal{D}) B - C_k \RB
\end{equation}
where $\text{Card}(\mathcal{D})$ is the number of \glspl{RE} carrying data and $\text{Card}(\mathcal{D}) M $ is the total number of bits transmitted by one user.
Then, following the derivations available in \cite{pilotless20}, $C_k$ can be written as 
\begin{align}
\label{eq:Ck}
\displaystyle
&C_k =  \sum_{(f, t)\in \mathcal{D}} \sum_{b=1}^{B} I\LB \mathscr{b}_{f,t,k,b}; \Ym \RB  \\
&-  \sum_{(f, t)\in \mathcal{D}} \sum_{b=1}^{B}  \EE_\Ym \LSB D_{KL} \LB P \LB  \mathscr{b}_{f,t,k,b} | \Ym \RB ||  \widetilde{P}_{\thetav} \LB  \mathscr{b}_{f,t,k,b} | \Ym \RB \RB \RSB \nonumber.
\end{align}
\hspace{-3pt}The first term in \eqref{eq:Ck} is the maximum information rate that can be achieved assuming an ideal \gls{BMD} receiver. 
It depends only on the transmitter and the channel, and therefore acts as a constant during the receiver training.
The second term in \eqref{eq:Ck} is the expected value of the KL-divergence between the true posterior probability $P \LB  \mathscr{b}_{f,t,k,m} | \Ym \RB$ and the one estimated by the proposed receiver.
The KL-divergence can be interpreted as a measure of distance between the true and estimated probabilities, which is minimized during training.
Moreover, $C_k$ is an achievable rate for user $k$ assuming a mismatched \gls{BMD} receiver \cite{pilotless20}, meaning that improvements in $C_k$ directly translate to an improved \gls{BER} performance.

\begin{figure}[b]

\centering
	
  \centering
  	\begin{subfigure}[b]{0.24\textwidth}
    	\begin{adjustbox}{height=0.7\linewidth} 
\begin{tikzpicture}

\begin{axis}[
colorbar,
colorbar style={ylabel={}},
colormap/viridis,
point meta max=0.1654548958409578,
point meta min=0.0362763414159417,
tick align=outside,
tick pos=left,
x grid style={white!69.0196078431373!black},
xmin=0.5, xmax=16.5,
xtick style={color=black},
y dir=reverse,
y grid style={white!69.0196078431373!black},
ymin=0.5, ymax=16.5,
ytick style={color=black},
width=7cm,height=7cm
]
\addplot graphics [includegraphics cmd=\pgfimage,xmin=0.5, xmax=16.5, ymin=16.5, ymax=0.5] {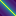};
\end{axis}

\end{tikzpicture}
  	\end{adjustbox}    
  	\caption{Amplitude}
	\end{subfigure}%
\hfill
  	\begin{subfigure}[b]{0.24\textwidth}
  	\begin{adjustbox}{height=0.7\linewidth} 
\begin{tikzpicture}

\begin{axis}[
colorbar,
colorbar style={ylabel={}},
colormap={mymap}{[1pt]
  rgb(0pt)=(0.2298057,0.298717966,0.753683153);
  rgb(1pt)=(0.26623388,0.353094838,0.801466763);
  rgb(2pt)=(0.30386891,0.406535296,0.84495867);
  rgb(3pt)=(0.342804478,0.458757618,0.883725899);
  rgb(4pt)=(0.38301334,0.50941904,0.917387822);
  rgb(5pt)=(0.424369608,0.558148092,0.945619588);
  rgb(6pt)=(0.46666708,0.604562568,0.968154911);
  rgb(7pt)=(0.509635204,0.648280772,0.98478814);
  rgb(8pt)=(0.552953156,0.688929332,0.995375608);
  rgb(9pt)=(0.596262162,0.726149107,0.999836203);
  rgb(10pt)=(0.639176211,0.759599947,0.998151185);
  rgb(11pt)=(0.681291281,0.788964712,0.990363227);
  rgb(12pt)=(0.722193294,0.813952739,0.976574709);
  rgb(13pt)=(0.761464949,0.834302879,0.956945269);
  rgb(14pt)=(0.798691636,0.849786142,0.931688648);
  rgb(15pt)=(0.833466556,0.860207984,0.901068838);
  rgb(16pt)=(0.865395197,0.86541021,0.865395561);
  rgb(17pt)=(0.897787179,0.848937047,0.820880546);
  rgb(18pt)=(0.924127593,0.827384882,0.774508472);
  rgb(19pt)=(0.944468518,0.800927443,0.726736146);
  rgb(20pt)=(0.958852946,0.769767752,0.678007945);
  rgb(21pt)=(0.96732803,0.734132809,0.628751763);
  rgb(22pt)=(0.969954137,0.694266682,0.579375448);
  rgb(23pt)=(0.966811177,0.650421156,0.530263762);
  rgb(24pt)=(0.958003065,0.602842431,0.481775914);
  rgb(25pt)=(0.943660866,0.551750968,0.434243684);
  rgb(26pt)=(0.923944917,0.49730856,0.387970225);
  rgb(27pt)=(0.89904617,0.439559467,0.343229596);
  rgb(28pt)=(0.869186849,0.378313092,0.300267182);
  rgb(29pt)=(0.834620542,0.312874446,0.259301199);
  rgb(30pt)=(0.795631745,0.24128379,0.220525627);
  rgb(31pt)=(0.752534934,0.157246067,0.184115123);
  rgb(32pt)=(0.705673158,0.01555616,0.150232812)
},
point meta max=2.44783252080282,
point meta min=-2.19349443912506,
tick align=outside,
tick pos=left,
x grid style={white!69.0196078431373!black},
xmin=0.5, xmax=16.5,
xtick style={color=black},
y dir=reverse,
y grid style={white!69.0196078431373!black},
ymin=0.5, ymax=16.5,
ytick style={color=black},
width=7cm,height=7cm
]
\addplot graphics [includegraphics cmd=\pgfimage,xmin=0.5, xmax=16.5, ymin=16.5, ymax=0.5] {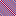};
\end{axis}

\end{tikzpicture}
  	\end{adjustbox} 
  	\caption{Phase}
	\end{subfigure}%

\caption{Example of amplitude and phase for $\Em_{f,t}^{(k)}$.}
\label{fig:E}
\end{figure}

\subsection{ML-enhanced channel estimator}

As seen in Section \ref{sec:stats}, the channel estimation error statistics can only be obtained for \glspl{RE} carying pilots.
However, the estimation accuracy decreases as we move away from them.
In the following, we present \glspl{CNN} that estimate the channel estimation error covariance matrices in the uplink and the estimation error variances in the downlink.

\subsubsection{Uplink scenario}
\label{sec:ml_ch_est_ul}

\begin{figure*}[t!]
    \centering
    \includegraphics[width=0.75\textwidth]{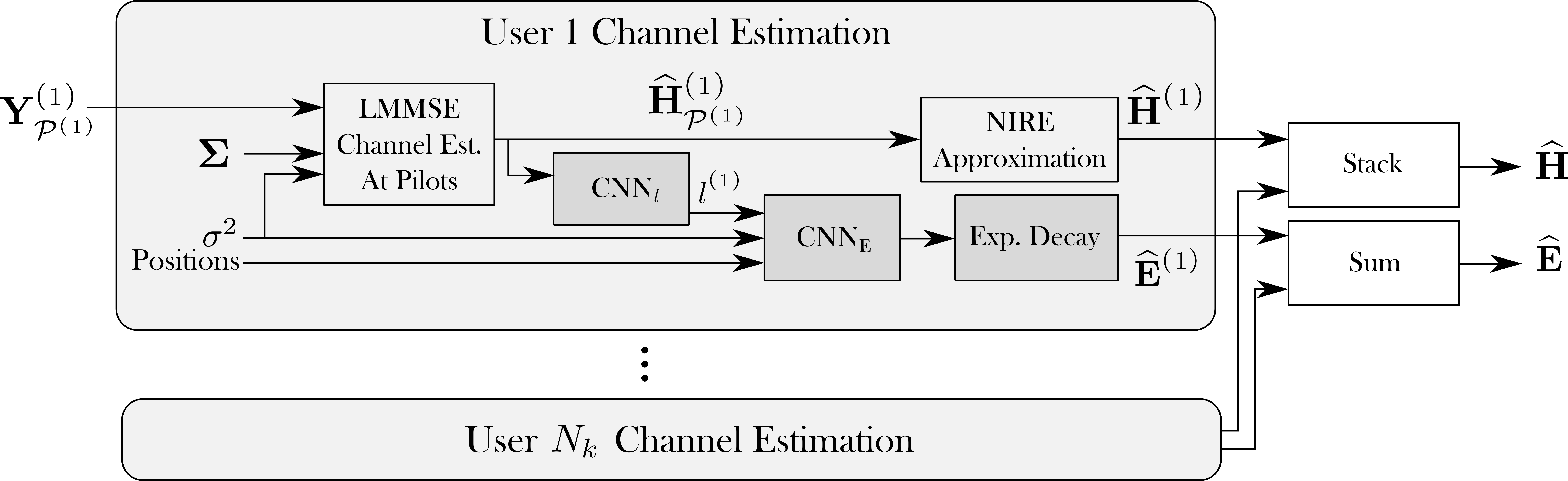}
    \caption{ML-enhanced uplink channel estimation.}
    \label{fig:ch_est_ml}
\end{figure*}

\begin{figure*}[b!]
    \centering
    \includegraphics[width=0.8\textwidth]{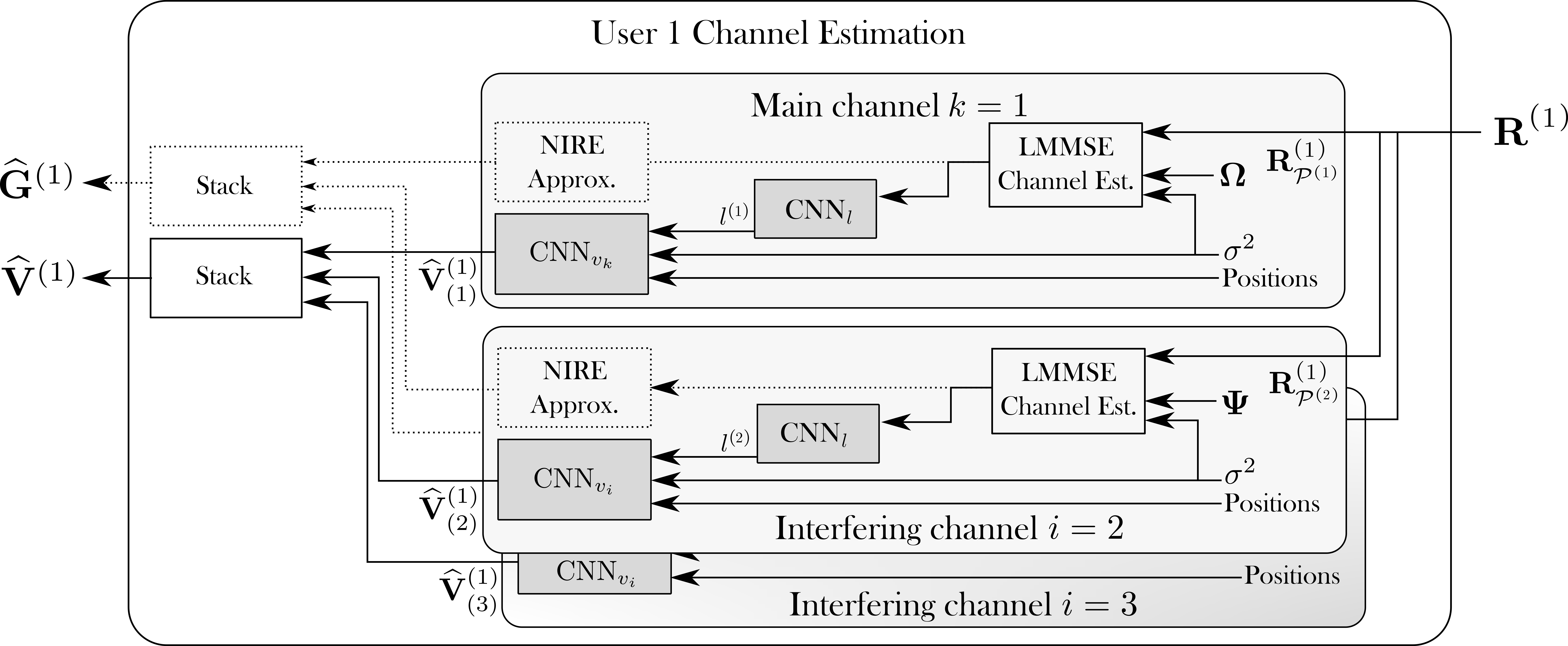}
    \caption{Detailed view of the downlink \emph{Channel Estimation} component of user $1$ out of $N_k = 3$, as depicted in Fig~\ref{fig:downlink}.}
    \label{fig:ch_est_dl}
\end{figure*}

The \gls{ML}-enhanced uplink channel estimation architecture is depicted in Fig.~\ref{fig:ch_est_ml}.
In this scenario, the spatial channel estimation error covariance matrices $\Em_{f,t}$ is needed to compute both the equalized symbols \eqref{eq:W} and the uplink post-equalization noise variance \eqref{eq:eq_snr_ul}.
An example of the amplitude and phase of an estimate of a covariance matrix $\widehat{\Em}_{1,1}^{(1)}$ is shown in Fig.~\ref{fig:E}  for a \gls{ULA} of antennas at the \gls{BS}.
One can see that the amplitude of the coefficients of $\widehat{\Em}^{(1)}_{1,1}$ decays rapidly when moving away from the diagonal.
The phase, on the other hand, exhibits a more surprising pattern, with a phase difference of roughly $\pi$ between two adjacent antennas.
To predict every element of $\widehat{\Em}^{(k)}$ for user $k$, a naively designed \gls{CNN} would need to output $N_f  N_t  N_m ^2$ complex parameters.
This would be of prohibitive complexity for any large number of subcarriers, symbols, or receiving antennas.
For this reason, we propose to approximate every element $(x,y)$ of $\widehat{\Em}_{f,t}^{(k)}$ with a complex power decay model:
\begin{equation}
\label{eq:exp_decay}
\hat{e}^{(k)}_{f,t,x,y}= \alpha_{f,t} \beta_{f,t}^{|y-x|} \exp{j \gamma (y-x)}
\end{equation}
where $y-x$ is the horizontal position difference between that element and the diagonal, and $\alpha_{f,t}$, $\beta_{f,t}$, and $\gamma$ are parameters of this model.
For a planar array, one could use such a model for each dimension, and take their Kronecker product to obtain the spatial channel estimation error covariance matrices.
A constant phase offset between two adjacent \glspl{RE} is assumed, which matches our experimental observations.
The parameters $\alpha_{f,t}$ and $\beta_{f,t}$ respectively control the scale and the decay of the model, and depend on the \gls{RE} $(f,t)$.

To estimate those two parameters for each \gls{RE}, we leverage a $\text{CNN}$, denoted by $\text{CNN}_{\mathbf{E}}$, which takes four inputs, each of size $N_f \times N_t$, for a total input dimension of $N_f \times N_t \times 4$.
$\text{CNN}_{\mathbf{E}}$ outputs $\alpha_{f,t}$ and $\beta_{f,t}$ for every \glspl{RE}, resulting in an output dimension of $N_f \times N_t \times 2$.
The first two inputs provide the location of every \glspl{RE} in the \gls{RG}.
More precisely, the first input matrix has all columns equal to $[-\frac{N_f}{2}, \dots, -1, 1, \dots, \frac{N_f}{2}]\tp$, whereas the second one has all rows equal to $[-\frac{N_t}{2}, \dots, -1, 1, \dots, \frac{N_t}{2}]$.
The third input provides the \gls{SNR} of the transmission and is given as a matrix $\text{SNR} \cdot \mathds{1}_{N_f \times N_t}$.
Finally, the fourth input is a feature $l^{(k)} \in \RR$ provided by another \gls{CNN}, denoted by $\text{CNN}_{l}$, which was designed with the intuition to predict the time-variability of the channel experienced by user $k$.
To do so, $\text{CNN}_{l}$ uses the channel estimates at \glspl{RE} carying pilots to estimate the Doppler and delay spread.
Although we cannot be certain that $\text{CNN}_l$ effectively learns to extract such information, the evaluations presented in Section \ref{sec:E_comp_speed} tend to support this hypothesis.
$\text{CNN}_{l}$ takes an input of dimension $N_{P_f} \times N_{P_t} \times 2N_m$, which corresponds to the stacking of the real and imaginary parts of $\widehat{\Hm}^{(k)}_{\mathcal{P}^{(k)}}$ along the last dimension.
It outputs the scalar $l^{(k)}$, which is fed to $\text{CNN}_{\mathbf{E}}$ as the matrix $l^{(k)}\cdot \mathds{1}_{N_f \times N_t}$.

\subsubsection{Downlink scenario}
\label{sec:ml_ch_est_dl}
In the downlink, the equivalent channel estimation error variances $\Vm^{(k)}$ are needed to compute $\tau^2_{f,t,k}$ in \eqref{eq:eq_snr_dl}.
To estimate those variances, we take inspiration from the architecture presented in Section~\ref{sec:system_model} which uses two different downlink covariance matrices $\Omegam$ and $\Psim$ to estimate the error variances of the main and interfering channels.
Similarly, we leverage two separate \glspl{CNN}, denoted by $\text{CNN}_{v_k}$ and by $\text{CNN}_{v_i}$, to respectively predict the estimation error variances of the main channel $\hat{v}^{(k)}_{f,t,k}$ and of an interfering channel $\hat{v}^{(k)}_{f,t,i}, i \neq k$.
Both \glspl{CNN} take the same inputs as $\text{CNN}_{\text{E}}$ but their outputs are of dimension $N_f \times N_t$ as variances are predicted for all \glspl{RE} $(f,t)$. 
To preserve the scalability of the conventional architecture, $N_k-1$ copies of $\text{CNN}_{v_i}$ are leveraged to estimate the error variances $\hat{v}^{(k)}_{f,t,i}$ of all $N_k - 1$ interferers.
The downlink channel estimation is schematically shown in Fig.~\ref{fig:ch_est_dl}, where $\widehat{\Vm}^{(k)}_{(a)} = \{ \hat{v}^{(k)}_{f,t,a} \}_{ (f,t) \in \{0, \dots, N_f \} \times \{0, \dots, N_t \}  }$ denotes the estimation error variances seen by user $k$ on its main or interfering channel $a$.

\begin{figure*}[t!]
  \centering
  \begin{subfigure}{1\textwidth}
    \includegraphics[width=1\textwidth]{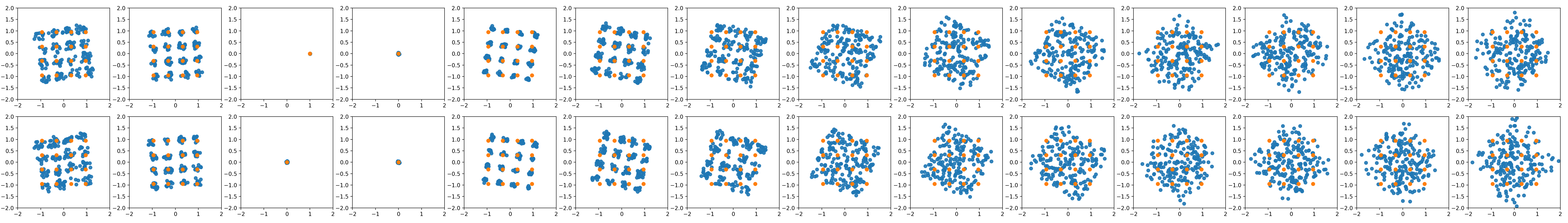}
    \caption{1P pilot pattern.}
    \label{fig:mismatch_1P}
    \vspace{10pt}
  \end{subfigure}
   \begin{subfigure}{1\textwidth}
    \includegraphics[width=1\textwidth]{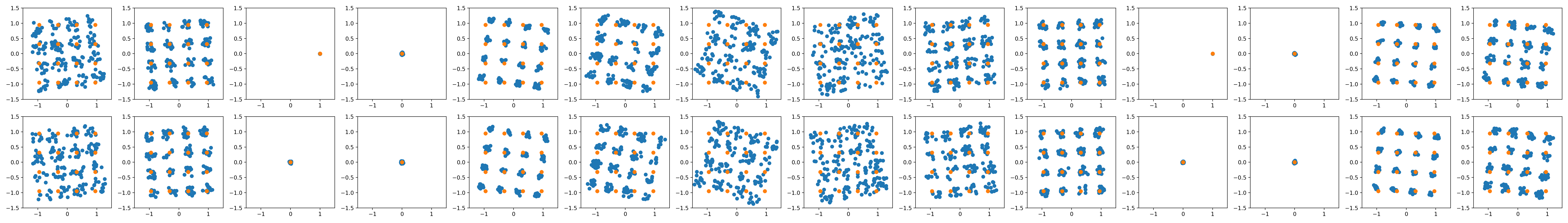}
    \caption{2P pilot pattern.}
    \label{fig:mismatch_2P}
  \end{subfigure}
  \caption{Mismatch between the sent signals (orange) and the equalized received ones (blue) for a single user out of four and using 16-QAM modulation.}
  \label{fig:mismatch}
\end{figure*}

\begin{figure*}[b!]
\centering
\begin{minipage}[b]{0.5\textwidth}
\centering
\includegraphics[height=90pt]{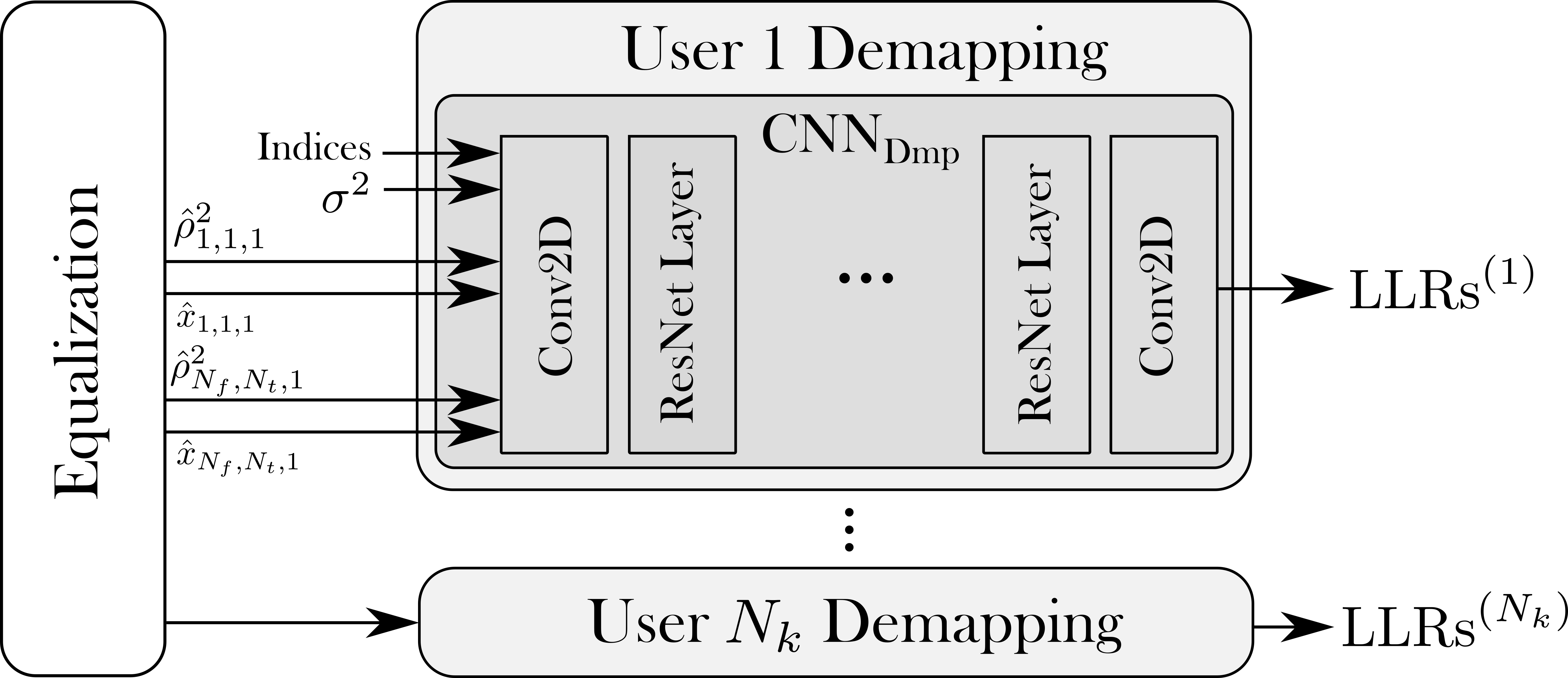}
  \captionof{figure}{Uplink demapping using the \gls{CNN}-based demapper.}
  \label{fig:demapper_ml}
\end{minipage}
\hfill
\begin{minipage}[b]{0.45\textwidth}
\centering
\includegraphics[height=90pt]{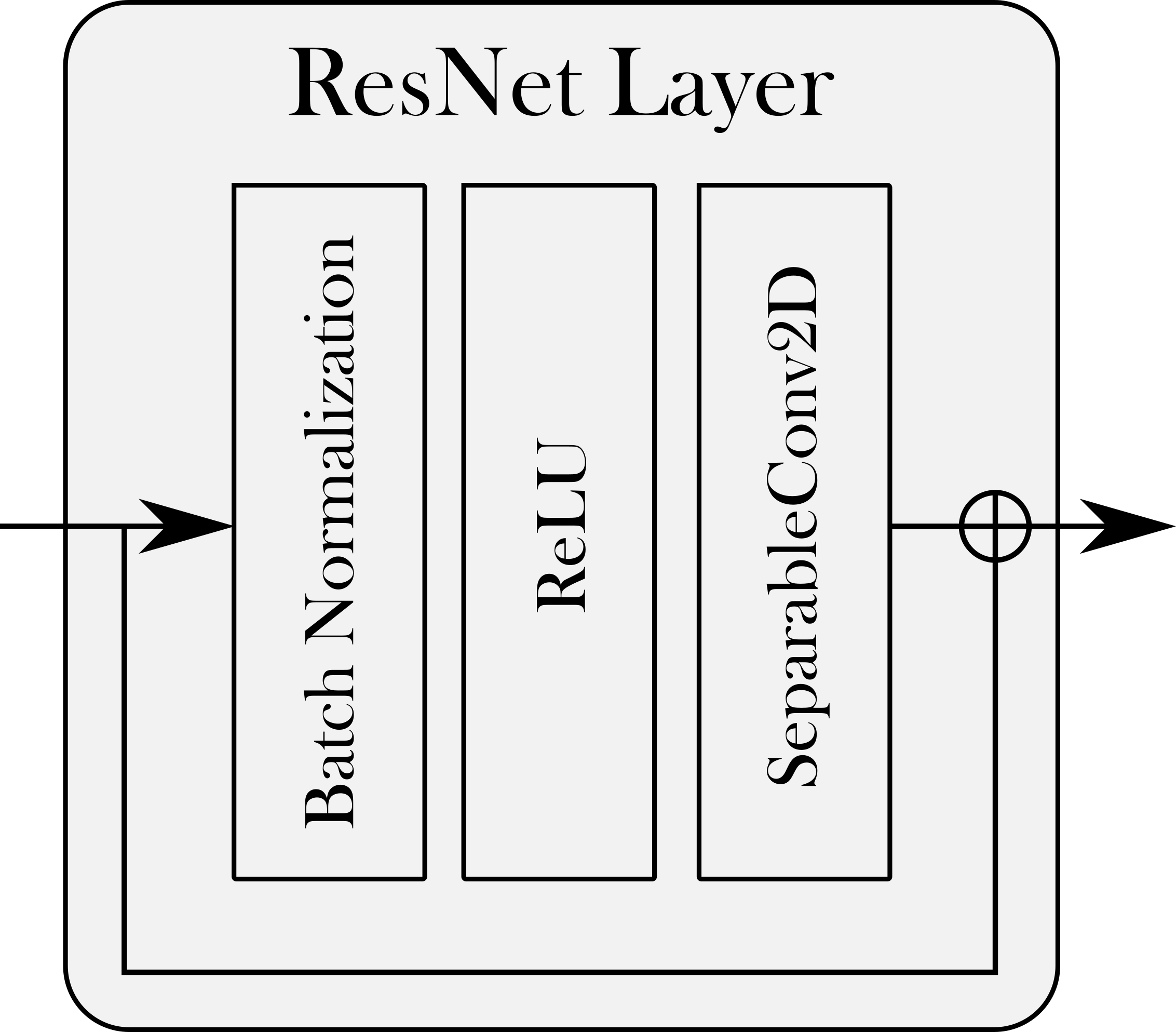}
  \captionof{figure}{Custom ResNet layer.}
  \label{fig:resnet}
\end{minipage}
\end{figure*}
\subsection{ML-enhanced demapper}
\label{sec:ml_demapper}

A consequence of unperfect channel estimation and equalization is channel aging, which leads to residual distortion on the equalized signals, as illustrated in Fig.\ref{fig:mismatch}. 
The signals sent by a single user out of four are represented in orange, while the corresponding equalized received signals are shown in blue, assuming spectral channel interpolation only. 
This figure has been obtained by sending a large batch of signals using a fixed realization of a fast-varying channel, and only displays the first two subcarriers of the uplink slot.
Moreover, an infinite \gls{SNR} is assumed so that only the effects of channel aging and interference are visible.
One can see that the equalized symbols suffer from little distortion and interference at \glspl{RE} close to the pilots, but these unwanted effects become increasingly stronger at \glspl{RE} that are away from them.

A traditional demapper, as presented in Section~\ref{sec:demapping_ul}, operates independently on each \gls{RE} and therefore only sees one equalized symbol at a time.
In contrast, we propose to use a \gls{CNN}, called $\text{CNN}_{\text{Dmp}}$, to perform a joint demapping of the entire \gls{RG}. 
By jointly processing all equalized symbols, the \gls{CNN} can estimate and correct the effects of channel aging to compute better \glspl{LLR}.
The input of $\text{CNN}_{\text{Dmp}}$ is of dimension $N_f \times N_t \times 6$ and carries the subcarriers and symbol indices for each RE, the SNR, the real and imaginary parts of the equalized symbols, and the post-equalization channel noise variances.
The output of $\text{CNN}_{\text{Dmp}}$ has dimensions $N_f \times N_t \times M$ and corresponds to the predicted $\text{LLRs}^{(k)}$ over the \gls{RG} for a user $k$. 
As with a conventional receiver, the demapping is performed independently for each user to make the architecture easily scalable.
The $\text{CNN}_{\text{Dmp}}$ demapper is shown in Fig.~\ref{fig:demapper_ml}, which depicts the uplink demapping process.

\begin{table*}
  \centering
  \renewcommand{\arraystretch}{1.5}
  \begin{tabular}{|p{1.8cm}||c|c|c|c||c|c|c|c||c|c|c|c|}
    \hline
     & \multicolumn{4}{c||}{$\text{CNN}_{\mathbf{E}} / \text{CNN}_{v_{k,i}}$} & \multicolumn{4}{c||}{$\text{CNN}_{l}$} & \multicolumn{4}{c|}{$\text{CNN}_{\text{Dmp}}$} \\ \hhline{|=||====||====||====|}
    Input size & \multicolumn{4}{c||}{$N_f \times N_k \times 4 $} & \multicolumn{4}{c||}{$N_{P_f} \times N_{P_t} \times 2N_m$} & \multicolumn{4}{c|}{$N_f \times N_k \times 6$} \\  \hline

    Parameters  & filters & kernel & dilat. & act. & filters & kernel & dilat. & act. & filters & kernel & dilat. & act. \\

    \hline
    Conv2D  & 32 & (5,3) & (1,1) & ReLU      & 32 & (1,1) & (1,1) & - & 128 & (1,1) & (1,1) & -  \\ \hline
    ResNet Layer  & \multicolumn{4}{c||}{-}  & 32 & (3,2) & (1,1) & - & 128 & (3,3) & (1,1) & -  \\ \hline
    ResNet Layer  & \multicolumn{4}{c||}{-}  & 32 & (5,2) & (2,1) & - & 128 & (5,3) & (2,1) & -  \\ \hline
    ResNet Layer  & \multicolumn{4}{c||}{-}  & 32 & (7,2) & (3,1) & - & 128 & (7,3) & (3,2) & -  \\ \hline
    ResNet Layer  & \multicolumn{4}{c||}{-}  & 32 & (5,2) & (2,1) & - & 128 & (9,3) & (4,3) & -  \\ \hline
    ResNet Layer  & \multicolumn{4}{c||}{-}  & 32 & (3,2) & (1,1) & - & 128 & (7,3) & (3,2) & -  \\ \hline
    ResNet Layer  & \multicolumn{4}{c||}{-}  & \multicolumn{4}{c||}{-} & 128 & (5,3) & (2,1) & -   \\ \hline
    ResNet Layer  & \multicolumn{4}{c||}{-}  & \multicolumn{4}{c||}{-} & 128 & (3,3) & (1,1) & -  \\ \hline
    Conv2D   & 32 & (5,3) & (1,1)& ReLU  & 1 &(3,2) & (1,1) & - & $M$  & (1,1) & (1,1) & - \\ \hline
    Conv2D   & 2 / 1 & (1,1) & (1,1)& Sigm.  & \multicolumn{4}{c||}{-}  & \multicolumn{4}{c|}{-}\\ \hline
    
    Output Layer & \multicolumn{4}{c||}{-} & \multicolumn{4}{c||}{Dense, units = 1}  & \multicolumn{4}{c|}{-} \\  \hline
    
    Output size  & \multicolumn{4}{c||}{$N_f \times N_k \times 2 / N_f \times N_k \times 1 $} & \multicolumn{4}{c||}{$1$} & \multicolumn{4}{c|}{$N_f \times N_k \times M$} \\  \hline
    
  \end{tabular}
  
  \caption{Parameters of the different CNNs.}
  \label{table:CNNs}
\end{table*}

\begin{figure*}[b!]
\centering
\begin{minipage}[b]{0.45\textwidth}
\centering
\includegraphics[height=70pt]{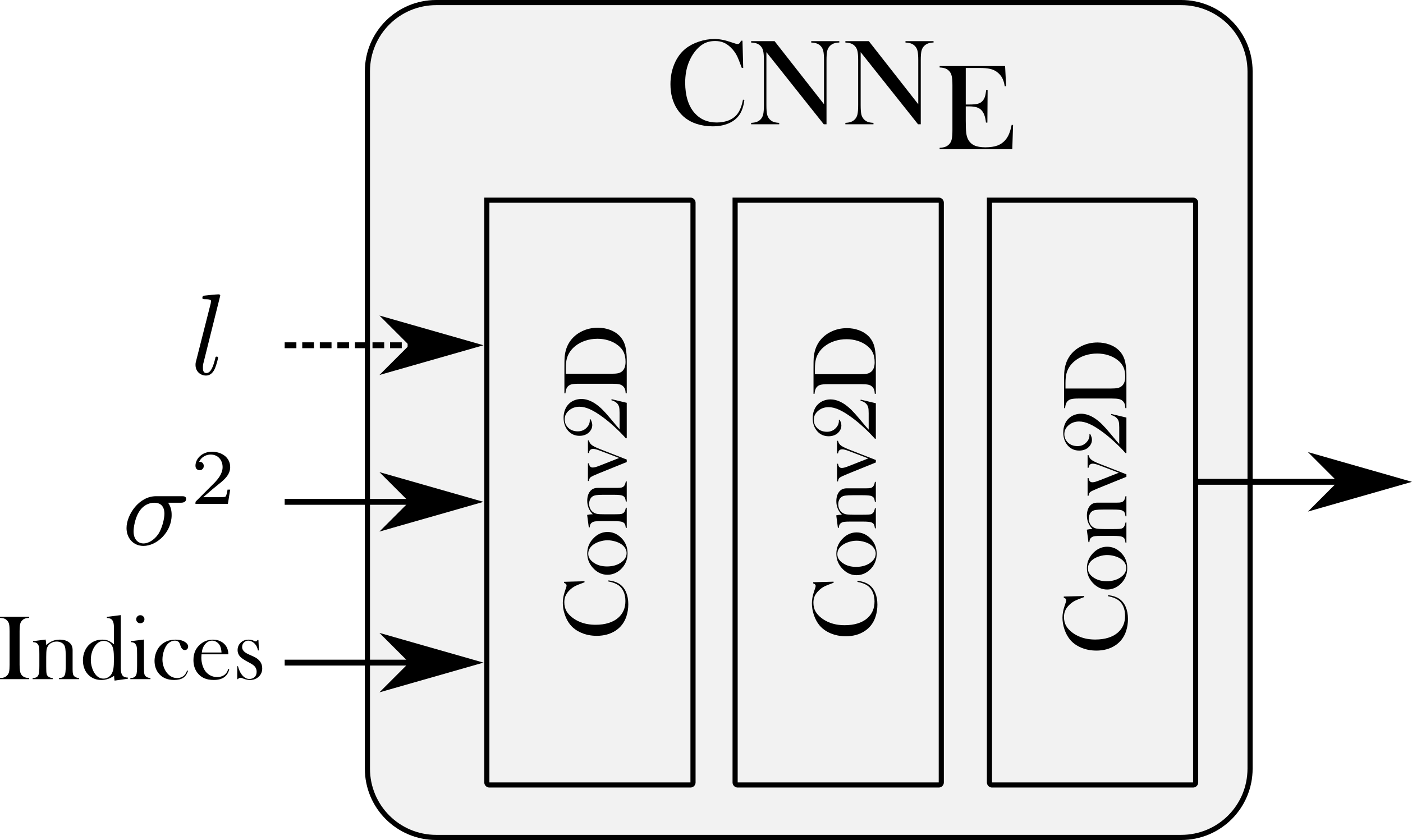}
  \captionof{figure}{Architecture of $\text{CNN}_{\mathbf{E}}$.}
  \label{fig:CNN_E}
\end{minipage}
\hfill
\begin{minipage}[b]{0.45\textwidth}
\centering
\includegraphics[height=70pt]{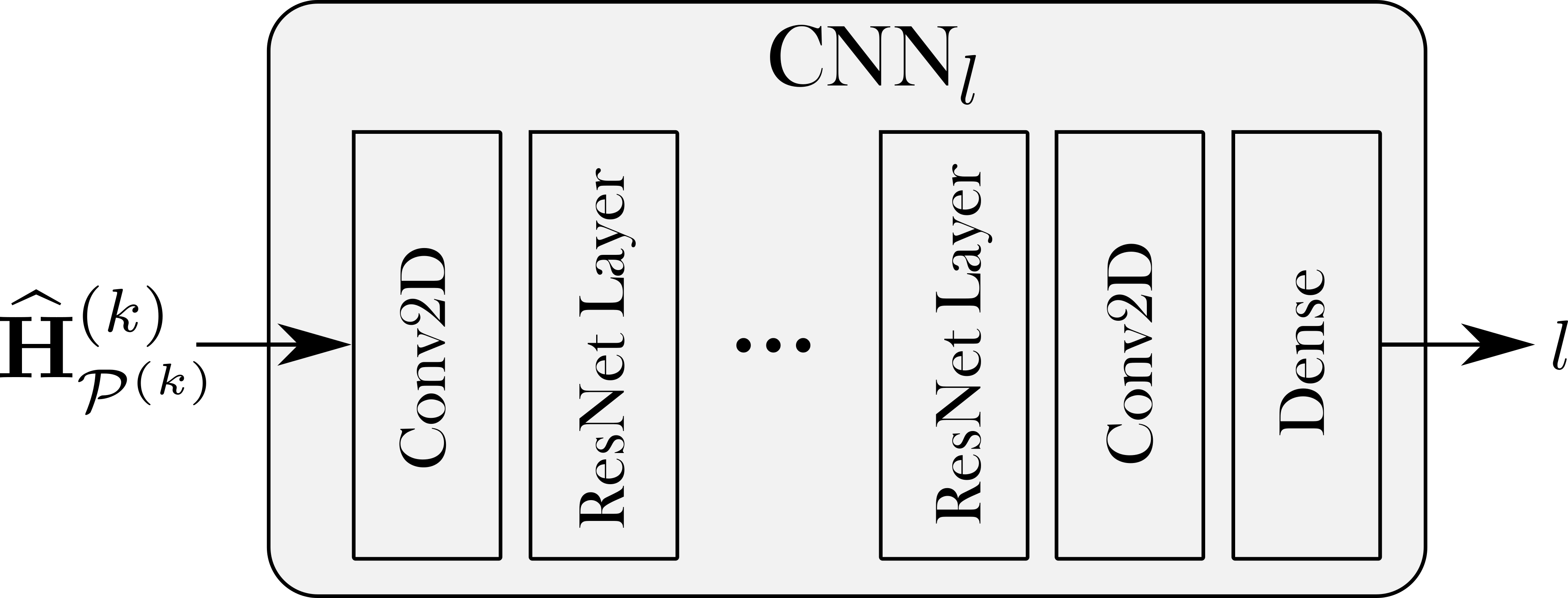}
  \captionof{figure}{Architecture of $\text{CNN}_{l}$.}
  \label{fig:CNN_l}
\end{minipage}
\end{figure*}

\subsection{CNN architectures}

All \glspl{CNN} presented above share the same building blocks: convolutional 2D layers, dense layers, and custom ResNet layers. 
The custom ResNet layers, inspired by \cite{he2016identity}, consist of a batch normalization layer, a \gls{ReLU}, a 2D separable convolutional layer, and finally the addition of the input, as depicted in Fig.~\ref{fig:resnet}.
Separable convolutions are less computationally expensive while maintaining similar performance as regular convolutional layers \cite{howard2017mobilenets}.
$\text{CNN}_{\mathbf{E}}$, $\text{CNN}_{v_k}$, and $\text{CNN}_{v_i}$ are all made of three 2D convolutional layers, as shown in Fig.\ref{fig:CNN_E} for $\text{CNN}_{\mathbf{E}}$. 
The first two layers are followed by a \gls{ReLU} activation function, while the last layer is followed by a sigmoid activation function.
$\text{CNN}_{\text{Dmp}}$ and $\text{CNN}_{l}$ also share a similar architecture, depicted in Fig.\ref{fig:demapper_ml}  and Fig~\ref{fig:CNN_l}, respectively.
Both are composed of a 2D convolutional layer, followed by multiple ResNet layers and a 2D convolutional layer. 
$\text{CNN}_{l}$ outputs a single scalar, which is ensured by using a dense layer with a single unit and no activation function as output layer.
Inspired by~\cite{honkala2020deeprx}, we used increasing followed by decreasing kernel sizes and dilation rates to increase the receptive field of the \gls{CNN}. 
All convolutional and separable convolutional layers use zero-padding so that the output dimensions matches the input dimensions.
Details of the architectures of all \glspl{CNN} are given in  Table~\ref{table:CNNs}.

\section{Evaluations} 
\label{sec:evaluations}

In this section, the proposed \gls{ML}-enhanced receiver is evaluated and compared against two baselines: the one presented in Section~\ref{sec:system_model} as well as a perfect \gls{CSI} baseline that will be detailed later on.
The training and evaluation setups are first introduced, followed by a numerical evaluation of the uplink and downlink performance for the 1P and 2P pilot patterns from Fig.~\ref{fig:channel_model}.

\begin{table*}[t!]
  \centering
  \renewcommand{\arraystretch}{1.5}
  \begin{tabular}{|p{4cm}|c|c|c|c|}
    \hline
    Parameters  & Symbol (if any)& \multicolumn{3}{c|}{Value}  \\ \hhline{|=|=|===|}
    Number of users & $N_k$ & \multicolumn{3}{c|}{4 (2 in high speed downlink)}   \\   \hline
    Number of antennas at the \gls{BS}& $N_m$ & \multicolumn{3}{c|}{16}  \\   \hline
    Number of subcarrier  & $N_f$ & \multicolumn{3}{c|}{72  \, = 6 resource blocks} \\   \hline
    Number of \gls{OFDM} symbols & $N_t$ & \multicolumn{3}{c|}{14 (uplink) + 14 (downlink)} \\   \hline
    Bit per channel use & $B$ & \multicolumn{3}{c|}{\SI{4}{\bit} (uplink), \SI{2}{\bit}(downlink)}  \\   \hline
    Center frequency & - & \multicolumn{3}{c|}{\SI{3.5}{\GHz}}  \\   \hline
    Subcarrier spacing & - & \multicolumn{3}{c|}{\SI{15}{\kHz}} \\   \hline
    Scenario & - & \multicolumn{3}{c|}{3GPP 38.901 UMi NLOS}  \\   \hline
    Code length & - & \multicolumn{3}{c|}{\SI{1296}{\bit}} \\   \hline
    Code rate & $\eta$ & \multicolumn{3}{c|}{$\frac{1}{2}$ (uplink), $\frac{1}{3}$ (downlink)} \\   \hline

    Learning rate & - & \multicolumn{3}{c|}{$10^{-3}$}  \\   \hline
    Batch size & $S$ & \multicolumn{3}{c|}{27 \glspl{RG}}   \\   \hline

  \end{tabular}
  
  \caption{Training and evaluation parameters.}
  \label{table:parameters}
\end{table*}

\subsection{Training and evaluation setup}

For realistic training and evaluation, the channel realizations were generated with QuaDRiGa version 2.0.0 \cite{quadriga}.
 It has been experimentally observed in \cite{honkala2020deeprx} that a receiver trained on certain scenarios was able to generalize to other channel models.
For this reason, we only focus on the \gls{3GPP} \gls{NLOS} \gls{UMi} scenario \cite{3gpp.38.901}.
The number of users was set to $N_k = 4$, except for the downlink at high speeds where it was reduced to $N_k=2$, and the number of antennas at the \gls{BS} was set to $N_m = 16$. 
All users were randomly placed within a \SI{120}{\degree} cell sector, with a minimum distance of \SI{15}{\meter} and a maximum distance of \SI{150}{\meter} from the \gls{BS}.
The user and \gls{BS} heights were respectively set to \SI{1.5}{\meter} and \SI{10}{\meter}.
The \glspl{RG} were composed of six resource blocks for a total of $N_f=72$ subarriers, with a center frequency of \SI{3.5}{\GHz} and a subcarrier spacing of \SI{15}{kHz}.
Both the uplink and downlink slots contained $N_t=14$ \gls{OFDM} symbols.
A Gray-labeled \gls{QAM} was used with $M= 4$ bits per channel use on the uplink and $M=2$ bits per channel use on the downlink.
The receivers were trained and evaluated on users moving at independent random speeds.
 Three ranges of low speeds were considered with the 1P pilot pattern : \SIrange[range-units=single]{0}{15}{\km\per\hour}, \SIrange[range-units=single]{15}{30}{\km\per\hour}, and \SIrange[range-units=single]{30}{45}{\km\per\hour}.
Similarly, three high speed ranges were considered with the 2P pilot pattern : \SIrange[range-units=single]{50}{70}{\km\per\hour}, \SIrange[range-units=single]{80}{100}{\km\per\hour}, and  \SIrange[range-units=single]{110}{130}{\km\per\hour}.
We noticed that $\text{CNN}_l$ was not able to extract useful information when using the 1P pattern, and therefore was not used in the corresponding trainings and evaluations.
For equalization, the grouped-\gls{LMMSE} equalizer operated on groups of $2\times 7$ \glspl{RE}, following the segmentation delimited by the thick black line of the 2P pattern, shown in Fig.~\ref{fig:2P_pattern}.
The training and evaluation parameters are provided in Table \ref{table:parameters}.

Separate training sets were constructed for the 1P and 2P pilot pattern, both made of channel realizations corresponding to 1000 \glspl{RG} of each respective user speed range, for a total of 3000 \glspl{RG} per training dataset.
Evaluations were performed separately for each user speed range with datasets containing 3000 \glspl{RG} and a standard IEEE 802.11n \gls{LDPC} code of length \SI{1296}{\bit} \cite{wlan}. Code rates of $\frac{1}{2}$ and $\frac{1}{3}$ were used respectively for uplink and downlink transmissions.
Decoding was done with 40 iterations using a conventional belief-propagation decoder.
To satisfy the perfect power assumption in \eqref{eq:snr}, each \gls{RG} was normalized to have an average energy of one per antenna and per user:
\begin{equation}
\sum_{f=1}^{N_f} \sum_{t=1}^{2N_t} ||\hv_{f,t}^{(k)}||_2^2 = N_f 2N_t N_m.
\end{equation} 

Each training was carried out using batches of size $S=27$ so that the total number of bits sent for each user was a multiple of the code length.
The trainable parameters $\thetav$ were all initialized randomly except for $\gamma$ in \eqref{eq:exp_decay} that was initialized with $\pi$.
Training was done through \gls{SGD} using the Adam \cite{Kingma15} optimizer and a learning rate of $10^{-3}$.
The best performing random initialization out of ten was selected.

\subsection{Uplink simulation results}

\begin{figure*}

\centering
	\begin{subfigure}[c]{1\textwidth}
	\centering
	\begin{tikzpicture} 
	
	\definecolor{color0}{rgb}{0.12156862745098,0.466666666666667,0.705882352941177}
	\definecolor{color1}{rgb}{1,0.498039215686275,0.0549019607843137}
	\definecolor{color2}{rgb}{0.172549019607843,0.627450980392157,0.172549019607843}
	\definecolor{color3}{rgb}{0.83921568627451,0.152941176470588,0.156862745098039}

    	\begin{axis}[%
    	hide axis,
    	xmin=10,
   	 xmax=50,
    	ymin=0,
    	ymax=0.4,
    	legend columns=5, 
    	legend style={draw=white!15!black,legend cell align=left,column sep=3.5ex}
    	]
    
    	\addlegendimage{white,mark=*, mark size=2}
    	\addlegendentry{\hspace{-1.25cm} Spectral interp. :};
    	\addlegendimage{color0,mark=*, mark size=2}
    	\addlegendentry{Baseline};
    	\addlegendimage{color1, mark=square*, mark size=1.5}
    	\addlegendentry{ML ch. est.};
    	\addlegendimage{color2, mark=diamond*, mark size=2}
    	\addlegendentry{ML receiver};
    	\addlegendimage{color3, mark=pentagon*, mark size=2}
    	\addlegendentry{Perfect CSI};
    	\end{axis}
	\end{tikzpicture}
	\end{subfigure}%
	\vspace{5pt}

  	\begin{subfigure}{0.4\textwidth}
  	
	\begin{tikzpicture} [trim left=3.57cm]
	
	\definecolor{color2}{rgb}{0.172549019607843,0.627450980392157,0.172549019607843}

    	\begin{axis}[%
    	hide axis,
    	xmin=10,
   	xmax=50,
    	ymin=0,
    	ymax=0.4,
    	legend columns=2, 
    	legend style={draw=white!15!black,legend cell align=right,column sep=0.5ex}
    	]
    	\addlegendimage{white,mark=square*, mark size=2}
    	\addlegendentry{\hspace{-0.9cm} 1P pattern :};
    	\addlegendimage{color2, dotted, mark=x, mark size=2, mark options={solid}}
    	\addlegendentry{\hspace{-0.05cm} ML receiver trained with $N_k=2$};
    	\end{axis}
	\end{tikzpicture}
	\vspace{0cm}
	\end{subfigure}
	

  	\begin{subfigure}{0.4\textwidth}
  	
	\vspace{-16.7pt}
	\begin{tikzpicture} [trim left=-5.2cm]
	
	\definecolor{color0}{rgb}{0.12156862745098,0.466666666666667,0.705882352941177}
	\definecolor{color2}{rgb}{0.172549019607843,0.627450980392157,0.172549019607843}

    	\begin{axis}[%
    	hide axis,
    	xmin=10,
   	 xmax=50,
    	ymin=0,
    	ymax=0.4,
    	legend columns=3, 
    	legend style={draw=white!15!black,legend cell align=left, column sep=0.5ex}
    	]
    	\addlegendimage{white,mark=square*, mark size=2}
    	\addlegendentry{\hspace{-0.9cm} 2P\,pattern,\,dual\,interp.\,:};
    	\addlegendimage{color0, dashed, mark=*, mark size=1, mark options={solid}}
    	\addlegendentry{\hspace{-0.15cm}  Baseline};
    	\addlegendimage{color2, dashed, mark=+, mark size=2, mark options={solid}}
    	\addlegendentry{\hspace{-0.15cm}  ML rec.};
    	\end{axis}
	\end{tikzpicture}
	\vspace{10pt}	
	\end{subfigure}%

  \centering
  	\begin{subfigure}[b]{0.32\textwidth}
  	\vspace{-10pt}
    	\begin{adjustbox}{width=\linewidth} 
\begin{tikzpicture}

\definecolor{color0}{rgb}{0.12156862745098,0.466666666666667,0.705882352941177}
\definecolor{color1}{rgb}{1,0.498039215686275,0.0549019607843137}
\definecolor{color2}{rgb}{0.172549019607843,0.627450980392157,0.172549019607843}
\definecolor{color3}{rgb}{0.83921568627451,0.152941176470588,0.156862745098039}

\begin{axis}[
log basis y={10},
tick align=outside,
tick pos=left,
x grid style={white!69.0196078431373!black},
xlabel={SNR},
xmajorgrids,
xmin=-5.75, xmax=10.75,
xtick style={color=black},
y grid style={white!69.0196078431373!black},
ylabel={BER},
ymajorgrids,
ymin=9e-06, ymax=0.284996815115009,
ymode=log,
ytick style={color=black}
]
\addplot [semithick, color0, mark=*, mark size=2, mark options={solid}]
table {%
-5 1.70455799e-01
-2.5 7.13982590e-02
0 1.61042529e-02
2.5 3.68813382e-03
5 8.11323304e-04
7.5 1.83531746e-04
10 3.74090606e-05
};

\addplot [semithick, color1, mark=square*, mark size=1.5, mark options={solid}]
table {%
-5 1.71270460e-01
-2.5 7.11075269e-02
0 1.61884806e-02
2.5 3.96611829e-03
5 8.65265370e-04
7.5 2.07085814e-04
10 3.68579153e-05
};

\addplot [semithick, color2, mark=diamond*, mark size=2, mark options={solid}]
table {%
-5 1.63974658e-01
-2.5 6.29884541e-02 
0.0 1.35914254e-02 
2.5 2.97567378e-03
5.0 5.10361550e-04 
7.5 1.17256394e-04 
10.0 1.54320987e-05
};

\addplot [semithick, dotted, color2, mark=x, mark size=2, mark options={solid}]
table {%
-5 0.16636697947978973
-2.5 0.06406939029693604
0 0.014092790834125011
2.5 0.0031420166441239418
5 0.0005471505739842542
7.5 0.00013356316078396048
10 2.0864197105984206e-05
};
 
\addplot [semithick, color3, mark=pentagon*, mark size=2, mark options={solid}]
table {%
-5 9.92080718e-02 
-2.5 2.60168653e-02 
0 3.39599513e-03 
2.5 5.64787257e-04
5 8.90790339e-05 
7.5 1.18736220e-05 
10 7.03339944e-07
};
\end{axis}

\end{tikzpicture}
  	\end{adjustbox}  
  	\vspace{-15pt}
  	\caption{1P pilot pattern at \SIrange[range-units=single]{0}{15}{\km\per\hour}.}
  	\label{fig:UL_1P_L}
	\end{subfigure}%
  	\hfill
  	\begin{subfigure}[b]{0.32\textwidth}
  	\vspace{-10pt}
    	\begin{adjustbox}{width=\linewidth} 
\begin{tikzpicture}

\definecolor{color0}{rgb}{0.12156862745098,0.466666666666667,0.705882352941177}
\definecolor{color1}{rgb}{1,0.498039215686275,0.0549019607843137}
\definecolor{color2}{rgb}{0.172549019607843,0.627450980392157,0.172549019607843}
\definecolor{color3}{rgb}{0.83921568627451,0.152941176470588,0.156862745098039}

\begin{axis}[
log basis y={10},
tick align=outside,
tick pos=left,
x grid style={white!69.0196078431373!black},
xlabel={SNR},
xmajorgrids,
xmin=-5.75, xmax=10.75,
xtick style={color=black},
y grid style={white!69.0196078431373!black},
ylabel={BER},
ymajorgrids,
ymin=4.45495867691486e-05, ymax=0.290400883476259,
ymode=log,
ytick style={color=black}
]
\addplot [semithick, color0, mark=*, mark size=2, mark options={solid}]
table {%
-5 0.1846478134393692
-2.5 0.08860826243956883
0 0.02830456556486232
2.5 0.008724582690774696
5 0.003121509410751363
7.5 0.0011042181134020212
10 0.0004941023669240529
};
\addplot [semithick, color1, mark=square*, mark size=1.5, mark options={solid}]
table {%
-5 0.1839657723903656
-2.5 0.08663561940193176
0 0.02595358162320086
2.5 0.007425405816320563
5 0.002154339366728285
7.5 0.0005230548226608274
10 0.0001574775866154232
};
\addplot [semithick, color2, mark=diamond*, mark size=2, mark options={solid}]
table {%
-5 0.1726982742547989
-2.5 0.07286844154198964
0 0.02001202671921679
2.5 0.00480035580767435
5 0.00123374117470424
7.5 0.00027827050336578395
10 8.65634173876606e-05
};

\addplot [semithick, dotted, color2, mark=x, mark size=2, mark options={solid}]
table {%
-5 0.1749269664287567
-2.5 0.07480434576670329
0 0.020850310434720347
2.5 0.005127921929670265
5 0.0013553608214715495
7.5 0.0003655563643202963
10 0.00011573243706958602
};

\addplot [semithick, color3, mark=pentagon*, mark size=2, mark options={solid}]
table {%
-5 0.11705315858125687
-2.5 0.037146118779977165
0 0.00776083000736045
2.5 0.001644020157527848
5 0.00036343511077575386
7.5 7.203740497866723e-05
10 2.1999927072708183e-05
};
\end{axis}

\end{tikzpicture}
  	\end{adjustbox}  
  	\vspace{-15pt}
  	\caption{1P pilot pattern at \SIrange[range-units=single]{15}{30}{\km\per\hour}.}  %
  	\label{fig:UL_1P_M}
	\end{subfigure}%
   \hfill
   \begin{subfigure}[b]{0.32\textwidth}
   \vspace{-10pt}
    	\begin{adjustbox}{width=\linewidth} 
\begin{tikzpicture}

\definecolor{color0}{rgb}{0.12156862745098,0.466666666666667,0.705882352941177}
\definecolor{color1}{rgb}{1,0.498039215686275,0.0549019607843137}
\definecolor{color2}{rgb}{0.172549019607843,0.627450980392157,0.172549019607843}
\definecolor{color3}{rgb}{0.83921568627451,0.152941176470588,0.156862745098039}

\begin{axis}[
log basis y={10},
tick align=outside,
tick pos=left,
x grid style={white!69.0196078431373!black},
xlabel={SNR},
xmajorgrids,
xmin=-5.75, xmax=10.75,
xtick style={color=black},
y grid style={white!69.0196078431373!black},
ylabel={BER},
ymajorgrids,
ymin=0.000204329461932803, ymax=0.296876644347008,
ymode=log,
ytick style={color=black}
]
\addplot [semithick, color0, mark=*, mark size=2, mark options={solid}]
table {%
-5 0.1975618600845337
-2.5 0.10643325746059418
0 0.06169325932860374
2.5 0.029828645288944244
5 0.01696120185910591
7.5 0.010197379318997264
10 0.008980654748156666
};
\addplot [semithick, color1, mark=square*, mark size=1.5, mark options={solid}]
table {%
-5 0.19705550372600555
-2.5 0.10215325653553009
0 0.05293554291129112
2.5 0.019646215485408902
5 0.008539406255641509
7.5 0.0032570684398524462
10 0.0014798969404910167
};
\addplot [semithick, color2, mark=diamond*, mark size=2, mark options={solid}]
table {%
-5 0.18415522575378418
-2.5 0.07863308861851692
0 0.031500082835555075
2.5 0.009625039558159187
5 0.003806682806628357
7.5 0.001328262786992127
10 0.0005091903651191388
};

\addplot [semithick, dotted, color2, mark=x, mark size=2, mark options={solid}]
table {%
-5 0.1855296492576599
-2.5 0.08241016417741776
0 0.03380377014875412
2.5 0.010749311887947845
5 0.004333014971197449
7.5 0.00171167318113148
10 0.0007236755946837365
};

\addplot [semithick, color3, mark=pentagon*, mark size=2, mark options={solid}]
table {%
-5 0.15017016232013702
-2.5 0.057083263993263245
0 0.022171241417527198
2.5 0.006593519210582599
5 0.002502959137061788
7.5 0.0007695381435769377
10 0.00039520943708950654
};
\end{axis}

\end{tikzpicture}
   		 
  	\end{adjustbox}    
  	\vspace{-15pt}
  	\caption{1P pilot pattern at \SIrange[range-units=single]{30}{45}{\km\per\hour}.}   %
  	\label{fig:UL_1P_H}
	\end{subfigure}%
   
   \vspace{10pt}
  	
  	\begin{subfigure}[b]{0.32\textwidth}
  	\begin{adjustbox}{width=\linewidth} 
\begin{tikzpicture}

\definecolor{color0}{rgb}{0.12156862745098,0.466666666666667,0.705882352941177}
\definecolor{color1}{rgb}{1,0.498039215686275,0.0549019607843137}
\definecolor{color2}{rgb}{0.172549019607843,0.627450980392157,0.172549019607843}
\definecolor{color3}{rgb}{0.83921568627451,0.152941176470588,0.156862745098039}

\begin{axis}[
log basis y={10},
tick align=outside,
tick pos=left,
x grid style={white!69.0196078431373!black},
xlabel={SNR},
xmajorgrids,
xmin=-5.75, xmax=10.75,
xtick style={color=black},
y grid style={white!69.0196078431373!black},
ylabel={BER},
ymajorgrids,
ymin=1.92917347126291e-05, ymax=0.296394799327145,
ymode=log,
ytick style={color=black}
]
\addplot [semithick, color0, mark=*, mark size=2, mark options={solid}]
table {%
-5 0.17254532873630524
-2.5 0.07584972977638245
0 0.02174021042883396
2.5 0.005923466421081685
5 0.0016146315616788344
7.5 0.0005304822564478674
10 0.00032807677547680214
};

\addplot [semithick, color1, mark=square*, mark size=1.5, mark options={solid}]
table {%
-5 0.17275028675794601
-2.5 0.07480227649211883
0 0.021221305802464487
2.5 0.005511332920286804
5 0.001409432866348652
7.5 0.0003266820994788578
10 0.00017727623516293532
};
\addplot [semithick, color2, mark=diamond*, mark size=2, mark options={solid}]
table {%
-5 0.1668788567185402
-2.5 0.06520061716437339
0 0.01731626146938652
2.5 0.004077305170940235
5 0.0009451195951260161
7.5 0.00023105517006712033
10 9.108410478394945e-05
};

\addplot [semithick, color3, mark=pentagon*, mark size=2, mark options={solid}]
table {%
-5 0.11548755690455437
-2.5 0.034257329627871515
0 0.007366174703929573
2.5 0.0013924575551209272
5 0.000289592979097506
7.5 7.14216814685642e-05
10 3.477044709143229e-05
};

\addplot [semithick, dashed, color0, mark=*, mark size=1, mark options={solid}]
table {%
-5 0.16387683898210526
-2.5 0.07110339589416981
0 0.0158540703356266
2.5 0.004082609954057261
5 0.0009156539328250801
7.5 0.00025631751583205187
10 0.00010894097156779026
};

\addplot [semithick, dashed, color2, mark=+, mark size=2, mark options={solid}]
table {%
-5 0.16032986342906952
-2.5 0.06646010279655457
0 0.014224054881681998
2.5 0.003549913188617211
5 0.0007263696057179913
7.5 0.0001927083330519963
10 8.029513934161514e-05
};

\end{axis}

\end{tikzpicture}
  	\end{adjustbox} 
  	\vspace{-15pt}
  	\caption{2P pilot pattern at \SIrange[range-units=single]{50}{70}{\km\per\hour}.}
  	\label{fig:UL_2P_L}
	\end{subfigure}%
	\hfill
  	\begin{subfigure}[b]{0.32\textwidth}
  	\begin{adjustbox}{width=\linewidth} 
\begin{tikzpicture}

\definecolor{color0}{rgb}{0.12156862745098,0.466666666666667,0.705882352941177}
\definecolor{color1}{rgb}{1,0.498039215686275,0.0549019607843137}
\definecolor{color2}{rgb}{0.172549019607843,0.627450980392157,0.172549019607843}
\definecolor{color3}{rgb}{0.83921568627451,0.152941176470588,0.156862745098039}

\begin{axis}[
log basis y={10},
tick align=outside,
tick pos=left,
x grid style={white!69.0196078431373!black},
xlabel={SNR},
xmajorgrids,
xmin=-5.75, xmax=10.75,
xtick style={color=black},
y grid style={white!69.0196078431373!black},
ylabel={BER},
ymajorgrids,
ymin=3e-5, ymax=0.29056388782237,
ymode=log,
ytick style={color=black}
]
\addplot [semithick, color0, mark=*, mark size=2, mark options={solid}]
table {%
-5 0.19766107201576233
-2.5 0.09440104166666667
0 0.04122154731303453
2.5 0.01689288727092472
5 0.006960069447134932
7.5 0.003392811199494948
10 0.002067515436404695
};

\addplot [semithick, color1, mark=square*, mark size=1.5, mark options={solid}]
table {%
-5 0.19739100337028503
-2.5 0.09111689527829488
0 0.03667872324585915
2.5 0.013129413438340029
5 0.003996141977647009
7.5 0.0013295396089840021
10 0.0004397505113956868
};
\addplot [semithick, color2, mark=pentagon*, mark size=2, mark options={solid}]
table {%
-5 0.18930844962596893
-2.5 0.07495659838120143
0 0.025046296417713165
2.5 0.007426112948451191
5 0.0018183513368542966
7.5 0.0005286265418462184
10 0.00013654578315249334
};

\addplot [semithick, color3, mark=diamond*, mark size=2, mark options={solid}]
table {%
-5 0.15181809663772583
-2.5 0.05091467499732971
0 0.015293209999799728
2.5 0.0044383358979371915
5 0.0010596707855196048
7.5 0.0002891589513455983
10 9.098508347960887e-05
};

\addplot [semithick, dashed, color0, mark=*, mark size=1, mark options={solid}]
table {%
-5 0.1812524050474167
-2.5 0.08607855997979641
0 0.025863500725891855
2.5 0.006777777799094717
5 0.0016710648123097296
7.5 0.00044164094332809324
10 0.00010615997982919605
};

\addplot [semithick, dashed, color2, mark=+, mark size=2, mark options={solid}]
table {%
-5 0.17500963807106018
-2.5 0.07563898526132107
0 0.021413376710067194
2.5 0.004997427999041975
5 0.000994997425202746
7.5 0.0002412808630935615
10 5.542695515032392e-05
};

\end{axis}
\end{tikzpicture}  		
  	\end{adjustbox} 
  	\vspace{-15pt}
  	\caption{2P pilot pattern at \SIrange[range-units=single]{80}{100}{\km\per\hour}.}
  	\label{fig:UL_2P_M}
	\end{subfigure}%
    \hfill
  	\begin{subfigure}[b]{0.32\textwidth}
  	\begin{adjustbox}{width=\linewidth} 
\begin{tikzpicture}

\definecolor{color0}{rgb}{0.12156862745098,0.466666666666667,0.705882352941177}
\definecolor{color1}{rgb}{1,0.498039215686275,0.0549019607843137}
\definecolor{color2}{rgb}{0.172549019607843,0.627450980392157,0.172549019607843}
\definecolor{color3}{rgb}{0.83921568627451,0.152941176470588,0.156862745098039}

\begin{axis}[
log basis y={10},
tick align=outside,
tick pos=left,
x grid style={white!69.0196078431373!black},
xlabel={SNR},
xmajorgrids,
xmin=-5.75, xmax=10.75,
xtick style={color=black},
y grid style={white!69.0196078431373!black},
ylabel={BER},
ymajorgrids,
ymin=0.000254659159228147, ymax=0.289449661539661,
ymode=log,
ytick style={color=black}
]
\addplot [semithick, color0, mark=*, mark size=2, mark options={solid}]
table {%
-5 0.21111111342906952
-2.5 0.13478250056505203
0 0.07800082117319107
2.5 0.0415351428091526
5 0.025429012477397917
7.5 0.01752727863419315
10 0.015883873384445905
};
\addplot [semithick, color1, mark=square*, mark size=1.5, mark options={solid}]
table {%
-5 0.21058546006679535
-2.5 0.1305362693965435
0 0.06325448386371135
2.5 0.027576108717105606
5 0.013051311802119017
7.5 0.006839816516730934
10 0.00412364972056821
};
\addplot [semithick, color2, mark=pentagon*, mark size=2, mark options={solid}]
table {%
-5 0.20206403732299805
-2.5 0.10108266025781631
0 0.03466493115574121
2.5 0.010000613729723475
5 0.0036591435558511877
7.5 0.001552121073156899
10 0.0007995756181480829
};
\addplot [semithick, color3, mark=diamond*, mark size=2, mark options={solid}]
table {%
-5 0.17277199029922485
-2.5 0.09070698171854019
0 0.038450183674693106
2.5 0.014035318343138153
5 0.0059697145223617555
7.5 0.0032043620683354043
10 0.0021673418187128845
};

\addplot [semithick, dashed, color0, mark=*, mark size=1, mark options={solid}]
table {%
-5 0.2037760466337204
-2.5 0.11408339689175288
0 0.041163676977157594
2.5 0.014065962368383622
5 0.005237557886168361
7.5 0.0019424768530327129
10 0.0008357445964429644
};

\addplot [semithick, dashed,  color2, mark=+, mark size=2, mark options={solid}]
table {%
-5 0.19689911603927612
-2.5 0.0950665498773257
0 0.030419078283011915
2.5 0.008801489799784927
5 0.002578800140647218
7.5 0.0007801118886767654
10 0.00034008487333267115
};

\end{axis}

\end{tikzpicture}
  	\end{adjustbox} 
  	\vspace{-15pt}
  	\caption{2P pilot pattern at \SIrange[range-units=single]{110}{130}{\km\per\hour}.}
  	\label{fig:UL_2P_H}
	\end{subfigure}%

\vspace{5pt}
\caption{Uplink BER achieved by the different receivers with the 1P and 2P pilot patterns.}
\vspace{0pt}
\label{fig:eval_UL}
\end{figure*}
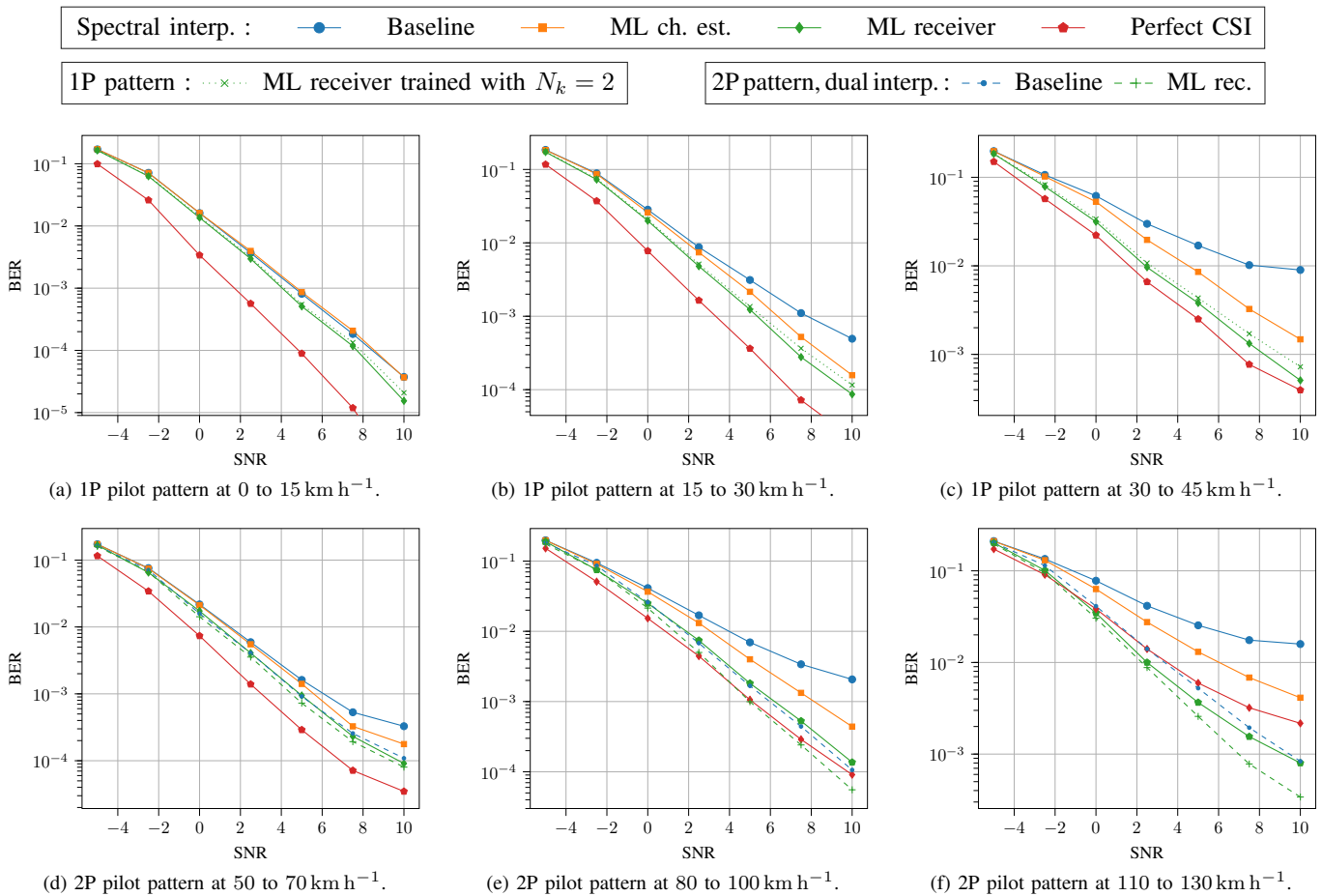

Different schemes were benchmarked in the uplink simulations.
The first one is the uplink baseline presented in  Sections~\ref{sec:baseline_ul} and~\ref{sec:stats}.
The second one, named `ML channel estimator', leverages the ML-enhanced channel estimator presented in \ref{sec:ml_ch_est_ul} but  is trained and tested with a standard demapper.
The third one is the \gls{ML}-enhanced receiver, leveraging both the enhanced channel estimator and  demapper.
We refer to it as `\gls{ML} receiver'.
Evaluating the \gls{ML} channel estimator separately allows us to better understand the role both components play in the observed gains.
An ideal baseline with perfect knowledge of the channel at the \glspl{RE} carrying pilots and of $\Em$ is also considered, and referred to as `Perfect \gls{CSI}'.
All schemes used spectral interpolation followed by \gls{NIRE} approximation for channel estimation.
Additional simulations were conducted for the 2P pilot pattern using spectral and temporal interpolation for both the baseline and the ML receiver.
Finally, an ML receiver trained with only $N_k=2$ users was also evaluated.

Simulation results for the 1P pattern are shown in the first row of Fig.~\ref{fig:eval_UL} for the three different speed ranges.
At speeds ranging from \SIrange[range-units=single]{0}{15}{\km\per\hour}, one can see that the \gls{ML} channel estimator does not bring any improvements, whereas the \gls{ML} receiver achieves a \SI{1}{\dB} gain at a coded \gls{BER} of $10^{-3}$.
The benefit of having a better estimation of $\Em$  becomes more significant as the speed increases.
At the highest speed range (Fig.~\ref{fig:UL_1P_H}), the \gls{ML} receiver enables gains of \SI{3}{\dB} over the baseline at a coded \gls{BER} of $10^{-2}$.
It can be observed that the ML receiver trained with only two users nearly matches the performance of the one trained with four users on all considered speeds, demonstrating the scalability of the proposed scheme with respect to the number of users.

Results for the 2P pilot pattern are shown in the second row of Fig.~\ref{fig:eval_UL}.
The gains provided by the \gls{ML} channel estimator alone and by the entire \gls{ML} receiver follow the same trend as with the 1P pattern, with moderate gains at \SI{50}{}-\SI{70}{\km\per\hour}, but significant improvements at \SI{110}{}-\SI{130}{\km\per\hour}.
More precisely, the \gls{ML} receiver provides a \SI{1}{\dB} gain over the baseline at a coded \gls{BER} of $10^{-3}$ in the \SI{50}{}-\SI{70}{\km\per\hour} range, and is the only scheme that achieves a coded \gls{BER} of $10^{-3}$ for the highest speeds with spectral interpolation only.
Indeed, at high speeds, the learned demapper is still able to mitigate the effects of channel aging, whereas even the perfect \gls{CSI} baseline suffers from strongly distorded equalized signals.
Using both spectral and temporal interpolations reduces the gains provided by the ML receiver, which can be explained by the better channel estimates leading to less channel aging, but they still amount to a \SI{2.2}{\dB} gap at a \gls{BER} of $10^{-3}$ for the highest speeds.
We have also experimentally verified that an ML receiver trained with only $N_k=2$ users was able to closely match the performance of the one trained with $N_k=4$, but decided not to include the corresponding curves for clarity reasons.
Overall, one can see that only the combination of a \gls{CNN}-based estimation of the channel estimation error statistics and \gls{CNN}-based demapper enables gains for both pilot patterns and all speed ranges.

\subsection{Visualizing the channel estimation error statistics}
\label{sec:E_comp_speed}

\begin{figure*}[t!]
\centering
\begin{minipage}[b]{0.55\textwidth}
  	\begin{subfigure}[b]{0.4\textwidth}
  	\begin{adjustbox}{height=6cm} 
\begin{tikzpicture}

\begin{groupplot}[group style={group size=2 by 1, 
								horizontal sep = 1 cm},
								width = 3.5cm,
								height = 9cm]
\nextgroupplot[
tick align=outside,
tick pos=left,
title={Predicted},
x grid style={white!69.0196078431373!black},
xmin=0.5, xmax=14.5,
xtick style={color=black},
y dir=reverse,
y grid style={white!69.0196078431373!black},
ymin=0.5, ymax=72.5,
ytick style={color=black}
]
\addplot graphics [includegraphics cmd=\pgfimage,xmin=0.5, xmax=14.5, ymin=72.5, ymax=0.5] {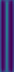};

\nextgroupplot[
tick align=outside,
tick pos=left,
title={True},
x grid style={white!69.0196078431373!black},
xmin=0.5, xmax=14.5,
xtick style={color=black},
y dir=reverse,
y grid style={white!69.0196078431373!black},
ymin=0.5, ymax=72.5,
ytick style={color=black}
]
\addplot graphics [includegraphics cmd=\pgfimage,xmin=0.5, xmax=14.5, ymin=72.5, ymax=0.5] {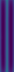};
\end{groupplot}

\end{tikzpicture}
  	\end{adjustbox} 
  	\caption{\SI{60}{\km\per\hour}}
	\end{subfigure}%
\hfill
  	\begin{subfigure}[b]{0.50\textwidth}
  	\begin{adjustbox}{height=6cm} 
\begin{tikzpicture}

\begin{groupplot}[group style={group size=2 by 1, 
								horizontal sep = 1 cm},
								width = 3.5cm,
								height = 9cm]
\nextgroupplot[
tick align=outside,
tick pos=left,
title={Predicted},
x grid style={white!69.0196078431373!black},
xmin=0.5, xmax=14.5,
xtick style={color=black},
y dir=reverse,
y grid style={white!69.0196078431373!black},
ymin=0.5, ymax=72.5,
ytick style={color=black}
]
\addplot graphics [includegraphics cmd=\pgfimage,xmin=0.5, xmax=14.5, ymin=72.5, ymax=0.5] {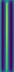};

\nextgroupplot[
colorbar,
colorbar style={ylabel={}},
colormap/viridis,
point meta max=1.2145604,
point meta min=0.11062375,
tick align=outside,
tick pos=left,
title={True},
x grid style={white!69.0196078431373!black},
xmin=0.5, xmax=14.5,
xtick style={color=black},
y dir=reverse,
y grid style={white!69.0196078431373!black},
ymin=0.5, ymax=72.5,
ytick style={color=black}
]
\addplot graphics [includegraphics cmd=\pgfimage,xmin=0.5, xmax=14.5, ymin=72.5, ymax=0.5] {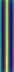};

\end{groupplot}

\end{tikzpicture}
  	\end{adjustbox} 
  	\caption{\SI{120}{\km\per\hour}}  	
  	\label{fig:E_RG_120}
	\end{subfigure}%
\caption{$||\widehat{\Em}_{f,t}||_{\text{F}}$ vs $||\Em_{f,t}||_{\text{F}}$ for different user speeds.}
\label{fig:E_RG}
\end{minipage}
\hspace{35pt}
\begin{minipage}[b]{0.30\textwidth}
\hfill
  	\begin{subfigure}[b]{0.40\textwidth}
  	\begin{adjustbox}{height=6cm} 
\begin{tikzpicture}

\begin{groupplot}[group style={group size=1 by 1, 
								horizontal sep = 1 cm},
								width = 3.5cm,
								height = 9cm]
\nextgroupplot[
tick align=outside,
tick pos=left,
title={Error},
title style = {color = white} ,
x grid style={white!69.0196078431373!black},
xmin=0.5, xmax=14.5,
xtick style={color=black},
y dir=reverse,
y grid style={white!69.0196078431373!black},
ymin=0.5, ymax=72.5,
ytick style={color=black}
]
\addplot graphics [includegraphics cmd=\pgfimage,xmin=0.5, xmax=14.5, ymin=72.5, ymax=0.5] {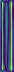};
\end{groupplot}

\end{tikzpicture}
  	\end{adjustbox} 
  	\caption{\SI{60}{\km\per\hour}}
	\end{subfigure}%
\hfill
  	\begin{subfigure}[b]{0.50\textwidth}
  	\begin{adjustbox}{height=6cm} 
\begin{tikzpicture}

\begin{groupplot}[group style={group size=1 by 1, 
								horizontal sep = 1 cm},
								width = 3.5cm,
								height = 9cm]
\nextgroupplot[
colorbar,
colorbar style={ylabel={}},
colormap/viridis,
point meta max=0.75,
point meta min=3.75502822862472e-05,
tick align=outside,
tick pos=left,
title={Error},
title style = {color = white} ,
x grid style={white!69.0196078431373!black},
xmin=0.5, xmax=14.5,
xtick style={color=black},
y dir=reverse,
y grid style={white!69.0196078431373!black},
ymin=0.5, ymax=72.5,
ytick style={color=black}
]
\addplot graphics [includegraphics cmd=\pgfimage,xmin=0.5, xmax=14.5, ymin=72.5, ymax=0.5] {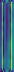};
\end{groupplot}

\end{tikzpicture}
  	\end{adjustbox} 
  	\caption{\SI{120}{\km\per\hour}}  	
	\end{subfigure}%
\caption{Normalized error.}
\label{fig:E_errors}
\end{minipage}
\end{figure*}

\begin{figure*}[b!]

\centering
  	\begin{subfigure}[b]{0.49\textwidth}
  	\centering
    	\begin{adjustbox}{height=3.75cm} 
\begin{tikzpicture}

\begin{groupplot}[group style={group size=2 by 1, 
								horizontal sep = 1 cm},
								width = 4.7cm,
								height = 4.7cm]
\nextgroupplot[
tick align=outside,
tick pos=left,
title={Predicted},
x grid style={white!69.0196078431373!black},
xmin=0.5, xmax=16.5,
xtick style={color=black},
y dir=reverse,
y grid style={white!69.0196078431373!black},
ymin=0.5, ymax=16.5,
ytick style={color=black}
]
\addplot graphics [includegraphics cmd=\pgfimage,xmin=0.5, xmax=16.5, ymin=16.5, ymax=0.5] {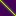};

\nextgroupplot[
colorbar,
colorbar style={ylabel={}},
colormap/viridis,
point meta max=0.21383855,
point meta min=0,
tick align=outside,
tick pos=left,
title={True},
x grid style={white!69.0196078431373!black},
xmin=0.5, xmax=16.5,
xtick style={color=black},
y dir=reverse,
y grid style={white!69.0196078431373!black},
ymin=0.5, ymax=16.5,
ytick style={color=black}
]
\addplot graphics [includegraphics cmd=\pgfimage,xmin=0.5, xmax=16.5, ymin=16.5, ymax=0.5] {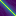};
\end{groupplot}

\end{tikzpicture}
  	\end{adjustbox}    
  	\caption{Amplitude}
  	\label{fig:E_RE_abs}
	\end{subfigure}%
\hfill
  	\begin{subfigure}[b]{0.49\textwidth}
  	\centering
  	\begin{adjustbox}{height=3.75cm} 
\begin{tikzpicture}

\begin{groupplot}[group style={group size=2 by 1, 
								horizontal sep = 1 cm},
								width = 4.7cm,
								height = 4.7cm]
\nextgroupplot[
tick align=outside,
tick pos=left,
title={Predicted},
x grid style={white!69.0196078431373!black},
xmin=0.5, xmax=16.5,
xtick style={color=black},
y dir=reverse,
y grid style={white!69.0196078431373!black},
ymin=0.5, ymax=16.5,
ytick style={color=black}
]
\addplot graphics [includegraphics cmd=\pgfimage,xmin=0.5, xmax=16.5, ymin=16.5, ymax=0.5] {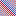};

\nextgroupplot[
colorbar,
colorbar style={ylabel={}},
colormap={mymap}{[1pt]
  rgb(0pt)=(0.2298057,0.298717966,0.753683153);
  rgb(1pt)=(0.26623388,0.353094838,0.801466763);
  rgb(2pt)=(0.30386891,0.406535296,0.84495867);
  rgb(3pt)=(0.342804478,0.458757618,0.883725899);
  rgb(4pt)=(0.38301334,0.50941904,0.917387822);
  rgb(5pt)=(0.424369608,0.558148092,0.945619588);
  rgb(6pt)=(0.46666708,0.604562568,0.968154911);
  rgb(7pt)=(0.509635204,0.648280772,0.98478814);
  rgb(8pt)=(0.552953156,0.688929332,0.995375608);
  rgb(9pt)=(0.596262162,0.726149107,0.999836203);
  rgb(10pt)=(0.639176211,0.759599947,0.998151185);
  rgb(11pt)=(0.681291281,0.788964712,0.990363227);
  rgb(12pt)=(0.722193294,0.813952739,0.976574709);
  rgb(13pt)=(0.761464949,0.834302879,0.956945269);
  rgb(14pt)=(0.798691636,0.849786142,0.931688648);
  rgb(15pt)=(0.833466556,0.860207984,0.901068838);
  rgb(16pt)=(0.865395197,0.86541021,0.865395561);
  rgb(17pt)=(0.897787179,0.848937047,0.820880546);
  rgb(18pt)=(0.924127593,0.827384882,0.774508472);
  rgb(19pt)=(0.944468518,0.800927443,0.726736146);
  rgb(20pt)=(0.958852946,0.769767752,0.678007945);
  rgb(21pt)=(0.96732803,0.734132809,0.628751763);
  rgb(22pt)=(0.969954137,0.694266682,0.579375448);
  rgb(23pt)=(0.966811177,0.650421156,0.530263762);
  rgb(24pt)=(0.958003065,0.602842431,0.481775914);
  rgb(25pt)=(0.943660866,0.551750968,0.434243684);
  rgb(26pt)=(0.923944917,0.49730856,0.387970225);
  rgb(27pt)=(0.89904617,0.439559467,0.343229596);
  rgb(28pt)=(0.869186849,0.378313092,0.300267182);
  rgb(29pt)=(0.834620542,0.312874446,0.259301199);
  rgb(30pt)=(0.795631745,0.24128379,0.220525627);
  rgb(31pt)=(0.752534934,0.157246067,0.184115123);
  rgb(32pt)=(0.705673158,0.01555616,0.150232812)
},
point meta max=2.91,
point meta min=-2.91,
tick align=outside,
tick pos=left,
title={True},
x grid style={white!69.0196078431373!black},
xmin=0.5, xmax=16.5,
xtick style={color=black},
y dir=reverse,
y grid style={white!69.0196078431373!black},
ymin=0.5, ymax=16.5,
ytick style={color=black}
]
\addplot graphics [includegraphics cmd=\pgfimage,xmin=0.5, xmax=16.5, ymin=16.5, ymax=0.5] {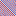};
\end{groupplot}

\end{tikzpicture}
  	\end{adjustbox} 
  	\caption{Phase}
  	\label{fig:E_RE_phase}
	\end{subfigure}%

\caption{Amplitude and phase of $\widehat{\Em}_{36, 7}$ and $\Em_{36, 7}$ for a single channel realization at \SI{120}{\km\per\hour}.}
\label{fig:E_RE}
\end{figure*}

In order to get insights into the \gls{ML} channel estimator abilities, we visualize the channel estimation error covariance matrices $\Em$ for different user speeds.
Two batches of uplink signals were sent, with users respectively moving at \SI{60}{\km\per\hour} and \SI{120}{\km\per\hour}.
All batches comprised 90 \glspl{RG}, and leveraged the 2P pilot pattern with an \gls{SNR} of \SI{5}{\dB}.
The Frobenius norms corresponding to each \glspl{RE}' $||\Em_{f,t}||_{\text{F}}$ are shown in Fig.~\ref{fig:E_RG}, and the normalized errors $\left| \frac{||\widehat{\Em}_{f,t}||_{\text{F}} - ||\Em_{f,t}||_{\text{F}}}{||\Em_{f,t}||_{\text{F}}} \right|$ are shown in Fig.~\ref{fig:E_errors}.
The figures labeled as `Predicted' refers to the estimations  $\widehat{\Em}_{f,t}$ produced by the \gls{ML} channel estimator and averaged over the corresponding batches, using spectral interpolation only.
As expected, the Frobenius norms of the \glspl{RE} carrying data strongly depend on their distance to their closest pilot, reflecting the distortions visible in  Fig.~\ref{fig:mismatch_2P}.
The figures labeled as `True' are presented for reference, and are computed by Monte Carlo simulations assuming knowledge of the true channel realizations.
One can see that the normalized errors are low on \glspl{RE} carrying data, confirming that $\text{CNN}_\Em$ is able to learn the channel estimation error statistics from the data during training.
The higher errors on \glspl{RE} carrying pilots are due to the loss \eqref{eq:loss_mc} taking solely into account the positions $(f,t)\in\mathcal{D}$, giving no opportunities for $\text{CNN}_\Em$ to learn the statistics at pilot positions.
Experiments conducted using both spectral and temporal interpolation yielded higher normalized errors, probably because the increased accuracy of the channel estimates results in a more difficult learning of the error statistics.
Additionally, $\text{CNN}_l$ seems to correctly estimate the time-variability of the channels since the Frobenius norms of the predicted covariances increase with the user speed, matching the behavior of the true covariances.

It is also insightful to look at the predicted and true spatial covariance matrix for a single \gls{RE}.
We chose to focus on the \gls{RE} $(36,7)$, positioned at the center of the \gls{OFDM} grid, and on the \SI{120}{\km\per\hour} scenario (Fig.~\ref{fig:E_RG_120}).
The amplitudes and phases of all elements of $\widehat{\Em}_{36, 7}$ and $\Em_{36, 7}$ are shown in Fig.~\ref{fig:E_RE_abs} and~\ref{fig:E_RE_phase}.
One can see that the predicted amplitudes and phases are close to the true ones, thus supporting the proposed power decay model.

\subsection{Downlink simulation results}

\begin{figure*}
\centering
	\begin{subfigure}[c]{1\textwidth}
	\centering
	\begin{tikzpicture} 
	
	\definecolor{color0}{rgb}{0.12156862745098,0.466666666666667,0.705882352941177}
	\definecolor{color1}{rgb}{1,0.498039215686275,0.0549019607843137}
	\definecolor{color2}{rgb}{0.172549019607843,0.627450980392157,0.172549019607843}
	\definecolor{color3}{rgb}{0.83921568627451,0.152941176470588,0.156862745098039}

    	\begin{axis}[%
    	hide axis,
    	xmin=10,
   	 xmax=50,
    	ymin=0,
    	ymax=0.4,
    	legend columns=5, 
    	legend style={draw=white!15!black,legend cell align=left,column sep=3.5ex}
    	]
    
    	\addlegendimage{white,mark=*, mark size=2}
    	\addlegendentry{\hspace{-1.25cm} Spectral interp. :};
    	\addlegendimage{color0,mark=*, mark size=2}
    	\addlegendentry{Baseline};
    	\addlegendimage{color1, mark=square*, mark size=1.5}
    	\addlegendentry{ML ch. est.};
    	\addlegendimage{color2, mark=diamond*, mark size=2}
    	\addlegendentry{ML receiver};
    	\addlegendimage{color3, mark=pentagon*, mark size=2}
    	\addlegendentry{Perfect CSI};
    	\end{axis}
	\end{tikzpicture}
	\end{subfigure}%
	\vspace{5pt}

  	\begin{subfigure}{0.4\textwidth}
  	
	\begin{tikzpicture} [trim left=3.57cm]
	
	\definecolor{color2}{rgb}{0.172549019607843,0.627450980392157,0.172549019607843}

    	\begin{axis}[%
    	hide axis,
    	xmin=10,
   	xmax=50,
    	ymin=0,
    	ymax=0.4,
    	legend columns=2, 
    	legend style={draw=white!15!black,legend cell align=right,column sep=0.5ex}
    	]
    	\addlegendimage{white,mark=square*, mark size=2}
    	\addlegendentry{\hspace{-0.9cm} 1P pattern :};
    	\addlegendimage{color2, dotted, mark=x, mark size=2, mark options={solid}}
    	\addlegendentry{\hspace{-0.05cm} ML receiver trained with $N_k=2$};
    	\end{axis}
	\end{tikzpicture}
	\vspace{0cm}
	\end{subfigure}
	

  	\begin{subfigure}{0.4\textwidth}
  	
	\vspace{-16.7pt}
	\begin{tikzpicture} [trim left=-5.2cm]
	
	\definecolor{color0}{rgb}{0.12156862745098,0.466666666666667,0.705882352941177}
	\definecolor{color2}{rgb}{0.172549019607843,0.627450980392157,0.172549019607843}

    	\begin{axis}[%
    	hide axis,
    	xmin=10,
   	 xmax=50,
    	ymin=0,
    	ymax=0.4,
    	legend columns=3, 
    	legend style={draw=white!15!black,legend cell align=left, column sep=0.5ex}
    	]
    	\addlegendimage{white,mark=square*, mark size=2}
    	\addlegendentry{\hspace{-0.9cm} 2P\,pattern,\,dual\,interp.\,:};
    	\addlegendimage{color0, dashed, mark=*, mark size=1, mark options={solid}}
    	\addlegendentry{\hspace{-0.15cm}  Baseline};
    	\addlegendimage{color2, dashed, mark=+, mark size=2, mark options={solid}}
    	\addlegendentry{\hspace{-0.15cm}  ML rec.};
    	\end{axis}
	\end{tikzpicture}
	\vspace{10pt}	
	\end{subfigure}%
	
  \centering
  	\begin{subfigure}[b]{0.32\textwidth}
  	\vspace{-10pt}
    	\begin{adjustbox}{width=\linewidth} 
\begin{tikzpicture}

\definecolor{color0}{rgb}{0.12156862745098,0.466666666666667,0.705882352941177}
\definecolor{color1}{rgb}{1,0.498039215686275,0.0549019607843137}
\definecolor{color2}{rgb}{0.172549019607843,0.627450980392157,0.172549019607843}
\definecolor{color3}{rgb}{0.83921568627451,0.152941176470588,0.156862745098039}

\begin{axis}[
log basis y={10},
tick align=outside,
tick pos=left,
x grid style={white!69.0196078431373!black},
xlabel={SNR},
xmajorgrids,
xmin=-5.5, xmax=5.5,
xtick style={color=black},
y grid style={white!69.0196078431373!black},
ylabel={BER},
ymajorgrids,
ymin=8e-6, ymax=0.95e-3,
ymode=log,
ytick style={color=black}
]
\addplot [semithick, color0, mark=*, mark size=2, mark options={solid}]
table {%
-5 0.0006372974575060653
-2.5 0.0001703042327426374
0 7.354662736062892e-05
2.5 1.777447065251181e-05
5 1.4224371893258648e-05
};
\addplot [semithick, color1, mark=square*, mark size=1.5, mark options={solid}]
table {%
-5 0.000564132775325561
-2.5 0.0001472594280494377
0 6.701223559910431e-05
2.5 1.5604331965732855e-05
5 1.129431203380227e-05
};
\addplot [semithick, color2, mark=diamond*, mark size=2, mark options={solid}]
table {%
-5 0.0005973048983560148
-2.5 0.0001336185514810495
0 5.6678242621710525e-05
2.5 1.653439152505598e-05
5 3.1436507826398957e-06
};
\addplot [semithick, color3, mark=pentagon*, mark size=2, mark options={solid}]
table {%
-5 0.0003048528461204114
-2.5 8.391203708015383e-05
0 4.091517767752521e-05
2.5 1.5087632054928691e-05
5 1.2500992303714155e-05
};

\addplot [semithick, dotted, color2, mark=x, mark size=2, mark options={solid}]
table {%
-5 0.0006994047629632405
-2.5 0.0001317584331263788
0 6.225859884987586e-05
2.5 1.9634590025816578e-05
5 3.573512811009721e-06
};
\end{axis}

\end{tikzpicture}
  	\end{adjustbox}  
  	\vspace{-15pt}  
  	\caption{1P pilot pattern, $N_k=4$, \SIrange[range-units=single]{0}{15}{\km\per\hour}.}
  	\label{fig:DL_1P_L}
	\end{subfigure}%
  	\hfill
  	\begin{subfigure}[b]{0.32\textwidth}
  	\vspace{-10pt}
    	\begin{adjustbox}{width=\linewidth} 
\begin{tikzpicture}

\definecolor{color0}{rgb}{0.12156862745098,0.466666666666667,0.705882352941177}
\definecolor{color1}{rgb}{1,0.498039215686275,0.0549019607843137}
\definecolor{color2}{rgb}{0.172549019607843,0.627450980392157,0.172549019607843}
\definecolor{color3}{rgb}{0.83921568627451,0.152941176470588,0.156862745098039}

\begin{axis}[
log basis y={10},
tick align=outside,
tick pos=left,
x grid style={white!69.0196078431373!black},
xlabel={SNR},
xmajorgrids,
xmin=-5.75, xmax=10.75,
xtick style={color=black},
y grid style={white!69.0196078431373!black},
ylabel={BER},
ymajorgrids,
ymin=4e-5, ymax=5e-3,
ymode=log,
ytick style={color=black}
]
\addplot [semithick, color0, mark=*, mark size=2, mark options={solid}]
table {%
-5 0.003484852652440572
-2.5 0.001374317953695936
0 0.0006828703688461246
2.5 0.00031632357929993303
5 0.0002652033731577831
7.5 0.00019386574213967834
10 0.00015376984306385565
};
\addplot [semithick, color1, mark=square*, mark size=1.5, mark options={solid}]
table {%
-5 0.0029258600867857845
-2.5 0.0008826264845993137
0 0.00044642856693826615
2.5 0.00017506530944141563
5 0.00014776868295390163
7.5 9.293402795563451e-05
10 8.409474230371415e-05
};
\addplot [semithick, color2, mark=diamond*, mark size=2, mark options={solid}]
table {%
-5 0.0025691458827673663
-2.5 0.0008026413695006341
0 0.00035042576072555673
2.5 0.00013609871159133034
5 8.881778969138395e-05
7.5 5.503224172163754e-05
10 5.089120452292264e-05
};
\addplot [semithick, color3, mark=pentagon*, mark size=2, mark options={solid}]
table {%
-5 0.0019299692196202362
-2.5 0.0007923073781239509
0 0.000436817953523132
2.5 0.00020461309403799533
5 0.0001664343618764542
7.5 0.00010074074047530302
10 9.74735444944963e-05
};

\addplot [semithick, dotted, color2, mark=x, mark size=2, mark options={solid}]
table {%
-5 0.00276204713143805
-2.5 0.0008273396154254442
0 0.00037367725140939
2.5 0.00015558201011619535
5 0.000112729000991676
7.5 7.202794338809326e-05
10 6.799768532800954e-05
};

\end{axis}

\end{tikzpicture}
  	\end{adjustbox}  
  	\vspace{-15pt}
  	\caption{1P pilot pattern, $N_k=4$, \SIrange[range-units=single]{15}{30}{\km\per\hour}.}  %
  	\label{fig:DL_1P_M}
	\end{subfigure}%
   \hfill
   \begin{subfigure}[b]{0.32\textwidth}
   \vspace{-10pt}
    	\begin{adjustbox}{width=\linewidth} 
\begin{tikzpicture}

\definecolor{color0}{rgb}{0.12156862745098,0.466666666666667,0.705882352941177}
\definecolor{color1}{rgb}{1,0.498039215686275,0.0549019607843137}
\definecolor{color2}{rgb}{0.172549019607843,0.627450980392157,0.172549019607843}
\definecolor{color3}{rgb}{0.83921568627451,0.152941176470588,0.156862745098039}

\begin{axis}[
log basis y={10},
tick align=outside,
tick pos=left,
x grid style={white!69.0196078431373!black},
xlabel={SNR},
xmajorgrids,
xmin=-5.75, xmax=10.75,
xtick style={color=black},
y grid style={white!69.0196078431373!black},
ylabel={BER},
ymajorgrids,
ymin=0.000842119810693929, ymax=3e-2,
ymode=log,
ytick style={color=black}
]
\addplot [semithick, color0, mark=*, mark size=2, mark options={solid}]
table {%
-5 0.020990410037338734
-2.5 0.011655092565342784
0 0.006659226190531626
2.5 0.0056545551757699285
5 0.0039275380151048015
7.5 0.0029714368391069003
10 0.0028729166946548507
};
\addplot [semithick, color1, mark=square*, mark size=1.5, mark options={solid}]
table {%
-5 0.017293733563274145
-2.5 0.00855303401593119
0 0.004114583348855377
2.5 0.0033992642082739622
5 0.0021459573544416344
7.5 0.001535218256758526
10 0.001522404106799513
};
\addplot [semithick, color2, mark=diamond*, mark size=2, mark options={solid}]
table {%
-5 0.015305266231298447
-2.5 0.007041997341439128
0 0.0033167989581124858
2.5 0.0025539434619713574
5 0.00144262567802798
7.5 0.0010428736785412184
10 0.0010251173834857766
};
\addplot [semithick, color3, mark=pentagon*, mark size=2, mark options={solid}]
table {%
-5 0.01423817795701325
-2.5 0.007917080027982592
0 0.004446717946557328
2.5 0.003995618398999797
5 0.0027955522475531323
7.5 0.0022852595872973323
10 0.002206101191313792
};

\addplot [semithick, color2, dotted, mark=x, mark size=2, mark options={solid}]
table {%
-5 0.016591435289010404
-2.5 0.007333002698142081
0 0.003521362461966783
2.5 0.00284680885153648
5 0.0017034556827275082
7.5 0.0012909226198098622
10 0.001267824074069358
};
\end{axis}

\end{tikzpicture}
  	\end{adjustbox}    
  	\vspace{-15pt}
  	\caption{1P pilot pattern, $N_k=4$, \SIrange[range-units=single]{30}{45}{\km\per\hour}.}   %
  	\label{fig:DL_1P_H}
	\end{subfigure}%
   
   \vspace{10pt}
  	
  	\begin{subfigure}[b]{0.32\textwidth}
  	\begin{adjustbox}{width=\linewidth} 
\begin{tikzpicture}

\definecolor{color0}{rgb}{0.12156862745098,0.466666666666667,0.705882352941177}
\definecolor{color1}{rgb}{1,0.498039215686275,0.0549019607843137}
\definecolor{color2}{rgb}{0.172549019607843,0.627450980392157,0.172549019607843}
\definecolor{color3}{rgb}{0.83921568627451,0.152941176470588,0.156862745098039}

\begin{axis}[
log basis y={10},
tick align=outside,
tick pos=left,
x grid style={white!69.0196078431373!black},
xlabel={SNR},
xmajorgrids,
xmin=-5.5, xmax=10.5,
xtick style={color=black},
y grid style={white!69.0196078431373!black},
ylabel={BER},
ymajorgrids,
ymin=0.00005, ymax=0.00991112757530391,
ymode=log,
ytick style={color=black}
]
\addplot [semithick, color0, mark=*, mark size=2, mark options={solid}]
table {%
-5 0.006798859211412492
-2.5 0.0028373016524710693
0 0.0014419643130531767
2.5 0.0009704563755402341
5 0.0006704067685799963
7.5 0.00048826389455483874
10 0.000498373028893111
};

\addplot [semithick, color1, mark=square*, mark size=1.5, mark options={solid}]
table {%
-5 0.0060086806358594915
-2.5 0.0022752976805350046
0 0.0011011904996121303
2.5 0.0006559524034149945
5 0.0003517857270082459
7.5 0.0002482242103661701
10 0.0002397718331810029
};

\addplot [semithick, color2, mark=diamond*, mark size=2, mark options={solid}]
table {%
-5 0.006256200466887094
-2.5 0.002485367135031993
0 0.001032490107063495
2.5 0.0005156547817024692
5 0.00023451389306690544
7.5 0.00014663690630579366
10 0.0001296230192342773
};

\addplot [semithick, color3, mark=pentagon*, mark size=2, mark options={solid}]
table {%
-5 0.004997023868563702
-2.5 0.0022152778258896434
0 0.0011277282029914205
2.5 0.0007809027990559115
5 0.0004888889049412683
7.5 0.0003541170682059601
10 0.00034571429656585673
};

\addplot [semithick, dashed, color0, mark=*, mark size=1, mark options={solid}]
table {%
-5 0.00628447425468039
-2.5 0.0020002480669063516
0 0.0008350694651744562
2.5 0.0004389881098177284
5 0.00023570436962880194
7.5 0.0001756944522378035
10 0.00012014682973947492
};

\addplot [semithick, dashed, color2, mark=+, mark size=2, mark options={solid}]
table {%
-5 0.00629464293437195
-2.5 0.0018725198843094404
0 0.0007123016094737977
2.5 0.0003606150882842485
5 0.00018021825809730216
7.5 0.00011133928614668547
10 7.000992132001557e-05
};

\end{axis}

\end{tikzpicture}
  	\end{adjustbox} 
  	\vspace{-15pt}
  	\caption{2P pilot pattern, $N_k=2$, \SIrange[range-units=single]{50}{70}{\km\per\hour}.}
  	\label{fig:DL_2P_L}
	\end{subfigure}%
	\hfill
  	\begin{subfigure}[b]{0.32\textwidth}
  	\begin{adjustbox}{width=\linewidth} 
\begin{tikzpicture}

\definecolor{color0}{rgb}{0.12156862745098,0.466666666666667,0.705882352941177}
\definecolor{color1}{rgb}{1,0.498039215686275,0.0549019607843137}
\definecolor{color2}{rgb}{0.172549019607843,0.627450980392157,0.172549019607843}
\definecolor{color3}{rgb}{0.83921568627451,0.152941176470588,0.156862745098039}

\begin{axis}[
log basis y={10},
tick align=outside,
tick pos=left,
x grid style={white!69.0196078431373!black},
xlabel={SNR},
xmajorgrids,
xmin=-5.5, xmax=10.5,
xtick style={color=black},
y grid style={white!69.0196078431373!black},
ylabel={BER},
ymajorgrids,
ymin=2.93e-4, ymax=4.11e-2,
ymode=log,
ytick style={color=black}
]
\addplot [semithick, color0, mark=*, mark size=2, mark options={solid}]
table {%
-5 0.029531249962747096
-2.5 0.011304563554003834
0 0.005731777415157443
2.5 0.004933261286527285
5 0.0034983292284538805
7.5 0.0032467532957153626
10 0.002754712349997135
};

\addplot [semithick, color1, mark=square*, mark size=1.5, mark options={solid}]
table {%
-5 0.027212797738611696
-2.5 0.009126488231122493
0 0.0041209795979962665
2.5 0.00309456175080303
5 0.0020885547629931303
7.5 0.0016109014737747862
10 0.0012348710597143509
};

\addplot [semithick, color2, mark=diamond*, mark size=2, mark options={solid}]
table {%
-5 0.027880456317216157
-2.5 0.009328373060561716
0 0.0034439397925764255
2.5 0.002286255458012888
5 0.0012808714833992877
7.5 0.0009046942865841794
10 0.0007111607300023025
};

\addplot [semithick, color3, mark=pentagon*, mark size=2, mark options={solid}]
table {%
-5 0.02419196425937116
-2.5 0.008829861213453114
0 0.004351590862170907
2.5 0.0037844968063643937
5 0.002739403260102508
7.5 0.0022409091375413647
10 0.0019408234456452195
};

\addplot [dashed, semithick, color0, mark=*, mark size=1, mark options={solid}]
table {%
-5 0.02205357117578387
-2.5 0.007842261992627754
0 0.0032932965458508543
2.5 0.001640724239619158
5 0.0011941964592551812
7.5 0.0008779762143967673
10 0.0007048611285063089
};

\addplot [dashed, semithick, color2, mark=+, mark size=2, mark options={solid}]
table {%
-5 0.022467261822894217
-2.5 0.008147817547433078
0 0.002817004850765777
2.5 0.0010080357317885501
5 0.0006495535897738592
7.5 0.00044751984865361007
10 0.0003847718289536715
};

\end{axis}

\end{tikzpicture}  		
  	\end{adjustbox} 
  	\vspace{-15pt}
  	\caption{2P pilot pattern, $N_k=2$, \SIrange[range-units=single]{80}{100}{\km\per\hour}.}
  	\label{fig:DL_2P_M}
	\end{subfigure}%
    \hfill
  	\begin{subfigure}[b]{0.32\textwidth}
  	\begin{adjustbox}{width=\linewidth} 
\begin{tikzpicture}

\definecolor{color0}{rgb}{0.12156862745098,0.466666666666667,0.705882352941177}
\definecolor{color1}{rgb}{1,0.498039215686275,0.0549019607843137}
\definecolor{color2}{rgb}{0.172549019607843,0.627450980392157,0.172549019607843}
\definecolor{color3}{rgb}{0.83921568627451,0.152941176470588,0.156862745098039}

\begin{axis}[
log basis y={10},
tick align=outside,
tick pos=left,
x grid style={white!69.0196078431373!black},
xlabel={SNR},
xmajorgrids,
xmin=-5.75, xmax=10.75,
xtick style={color=black},
y grid style={white!69.0196078431373!black},
ylabel={BER},
ymajorgrids,
ymin=6e-4, ymax=9e-2,
ymode=log,
ytick style={color=black}
]
\addplot [semithick, color0, mark=*, mark size=2, mark options={solid}]
table {%
-5 0.05973834302276373
-2.5 0.034306795429438354
0 0.017917906818911433
2.5 0.01172161189158662
5 0.008816588564183225
7.5 0.008333333427706235
10 0.008273605366401812
};

\addplot [semithick, color1, mark=square*, mark size=1.5, mark options={solid}]
table {%
-5 0.05558531749993563
-2.5 0.027022569393739104
0 0.01290178574854508
2.5 0.0067800673002806995
5 0.004680315125959389
7.5 0.003977640515408264
10 0.003430730642257806
};

\addplot [semithick, color2, mark=diamond*, mark size=2, mark options={solid}]
table {%
-5 0.05849826354533434
-2.5 0.026537698227912188
0 0.011266121128574013
2.5 0.005678762261773674
5 0.0038587362109260125
7.5 0.003048840104375417
10 0.0024119106441126753
};

\addplot [semithick, color3, mark=pentagon*, mark size=2, mark options={solid}]
table {%
-5 0.05361979175359011
-2.5 0.02692708345130086
0 0.01386904758401215
2.5 0.008893925645603585
5 0.006205026461432377
7.5 0.005404533062894375
10 0.0054696545848650305
};

\addplot [dashed, semithick, color0, mark=*, mark size=1, mark options={solid}]
table {%
-5 0.04173462312668562
-2.5 0.017502480121329426
0 0.008631944564403965
2.5 0.004616147797339811
5 0.003047295785738144
7.5 0.002501785768583377
10 0.002151587364378065
};

\addplot [dashed, semithick, color2, mark=+, mark size=2, mark options={solid}]
table {%
-5 0.043635912723839286
-2.5 0.017434027837589385
0 0.007866567526943981
2.5 0.0031078297054941335
5 0.0017131467108957162
7.5 0.001000770454393592
10 0.0008531746231892612
};

\end{axis}

\end{tikzpicture}
  	\end{adjustbox} 
  	\vspace{-15pt}
  	\caption{2P pilot pattern, $N_k=2$, \SIrange[range-units=single]{110}{130}{\km\per\hour}.}
  	\label{fig:DL_2P_H}
	\end{subfigure}%

\vspace{5pt}
\caption{Downlink BER achieved by the different receivers with the 1P and 2P pilot patterns.}
\vspace{-5pt}
\label{fig:eval_DL}
\end{figure*}
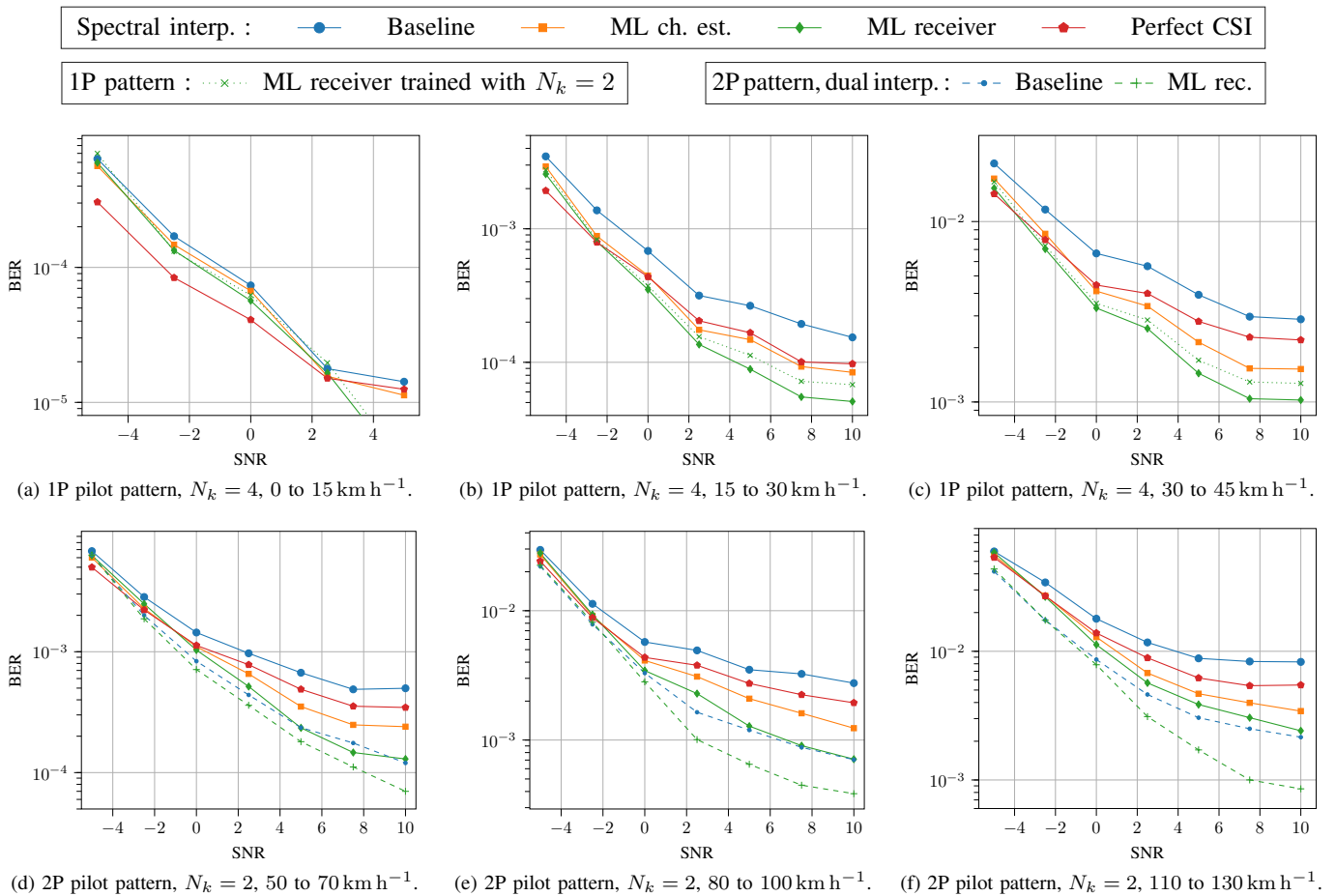

Four downlink schemes were evaluated on the considered speed ranges.
The first one is the baseline presented in Section~\ref{sec:system_model}.
In the second one, referred to as `\gls{ML} channel estimator', each user leverages the enhanced channel estimator of Section~\ref{sec:ml_ch_est_dl} and  is trained and tested with a standard demapper.
The third one is the \gls{ML}-enhanced receiver using both the enhanced channel estimator and \gls{ML} demapper for each user, and is referred to as `\gls{ML} receiver'.
The last scheme has perfect \gls{CSI}, i.e., all users perfectly know both the channel at \glspl{RE} carrying pilots and the estimation error variances $v^{(k)}_{f,t,k}$ and $v^{(k)}_{f,t,i}$ everywhere.
All schemes use the 2P pilot pattern to perform the uplink channel estimation, but are evaluated on both patterns in the downlink.
Similarly to the uplink, all schemes leveraged spectral interpolation with \gls{NIRE} approximation for channel estimation, but additional simulations were performed with the 2P pilot pattern using spectral and temporal interpolation for both the baseline and the ML receiver.
An ML receiver trained with only $N_k=2$ users was also evaluated on the three slowest speed ranges.

The 1P pilot pattern downlink evaluations are shown in the first row of Fig.~\ref{fig:eval_DL}.
Between \SI{0}{} and \SI{15}{\km\per\hour}, all schemes achieve good results but the ML receivers are the only ones to not saturate at high \glspl{SNR}.
As expected, the gains allowed by the learned channel estimator and demapper increase with the speed, and the \gls{ML} receiver is the only scheme to reach a \gls{BER} of approximately $10^{-3}$ in the \SI{30}{}-\SI{45}{\km\per\hour} speed range.
It can also be observed that the ML receiver trained with only $N_k=2$ users is able to closely match the performance of the ML receiver trained with four users.
The 2P pilot pattern evaluations are shown in the second row of Fig.~\ref{fig:eval_DL} for higher speeds.
In this new setup, the \gls{ML} receiver outperforms the baseline by \SI{1.1}{\dB} at a \gls{BER} of $10^{-3}$ in the speed range \SI{50}{}-\SI{70}{\km\per\hour}, and is the only one with spectral interpolation only to achieve a \gls{BER} of $10^{-3}$ in the speed range  \SI{70}{}-\SI{90}{\km\per\hour}.
Using both spectral and temporal interpolations, the ML receiver still provides significant gains at high speeds, being the only one to achieve a \gls{BER} of $10^{-3}$ in the \SI{110}{}-\SI{130}{\km\per\hour} speed range.

In all downlink scenarios except for the slowest speed range, the \gls{ML} channel estimator outperforms the perfect \gls{CSI} baseline.
This can be surprising since this baseline uses the exact noise variances $v^{(k)}_{f,t,k}$ and $v^{(k)}_{f,t,i}$ while the \gls{ML} schemes can only estimate them.
We suppose that this is because the baselines assume that the noise $o_{f,t,k}$ in \eqref{eq:eq_eq_ch_dl} is Gaussian distributed and uncorrelated to the transmitted signal, which is typically untrue.
The \gls{ML} channel estimator seems able to learn this model mismatch during training and predict variances that counteract it.

\section{Conclusion}
\label{sec:conclusion}

We propose in this paper a \gls{MU-MIMO} receiver that builds on top of a traditional receiver architecture and enhances it with \gls{ML} components.
More precisely, \glspl{CNN} are leveraged to improve both the demapping and the computation of the channel estimation error statistics.
All components of the proposed architecture are jointly optimized to refine the estimation of the \glspl{LLR}.
This approach does not require any knowledge of the channel for training, is interpretable, and easily scalable to any number of users.
Uplink and downlink evaluations were performed with multiple user speeds and two different pilot configurations on \gls{3GPP}-compliant channel models.
The results reveal that the proposed architecture effectively exploits the \gls{OFDM} structure to achieve tangible gains at low speeds and significant ones at high speeds compared to a traditional receiver.
On the one hand, the enhanced demapper jointly processes all \glspl{RE} of the \gls{OFDM} grid to counteract the effects of channel aging.
On the other hand, we demonstrated that the enhanced channel estimator is able to learn its error statistics during training.
In order to get insights into the improvements enabled by each of the trainable components, we also evaluated a conventional structure where only the channel estimator was enhanced.
The results indicate that the combined use of both the \gls{ML} channel estimator and demapper is key to achieve a substantial reduction of the coded \glspl{BER} across all scenarios.
Possible research directions include a more detailed study of the adaptability of the proposed architecture to any number of users, and enhancements to other parts of the receiver such as the equalization or the estimation of the channel coefficients.
We have also observed that further gains can be enabled by jointly processing all users as in \cite{korpi2020deeprx}.
However, unlocking these gains while preserving the flexibility and interpretability of conventional architectures is a significant challenge that is yet to be solved.

\appendix
\subsection{Grouped-LMMSE equalizer}
\label{app:LMMSE}

We aim to find the \gls{LMMSE} matrix to equalize a group of \glspl{RE} that spans multiple subcarriers $f \in [F_b, F_e]$ and symbols $t \in [T_b, T_e]$, with $1 \leq F_b \leq F_e \leq N_f$ and $1 \leq T_b \leq T_e \leq N_t$.
Let us denote by $\widehat{\Hm}_{f,t}$ the channel estimated at a \gls{RE} $(f,t)$ and by $\widetilde{\Hm}_{f,t}$ the corresponding estimation errors.
It is assumed that $\widehat{\Hm}$ is known, but that $\widetilde{\Hm}$, the symbols, and the noise, conditioned on the channel estimates, are random and uncorrelated.
The channel transfer function for that group is
\begin{equation}
\yv_{f, t} = \LB \widehat{\Hm}_{f,t} + \widetilde{\Hm}_{f,t} \RB \xv_{f,t} + \nv_{f,t}.
\end{equation}
We denote by $\Wm_{f,t}, \;f \in [F_b, F_e], t \in [T_b, T_e]$, the \gls{LMMSE} matrix that minimizes

\vspace{-5pt}
{\small
\begin{align}
& \mathcal{L}(\Wm_{f,t}) 
 =\sum_{f'=F_b}^{F_e} \sum_{t'=T_b}^{T_e} \text{MSE}\LB  \xv_{f',t'}, \Wm_{f,t} \yv_{f',t'} \RB \\
&= \EE \LSB \sum_{f'=F_b}^{F_e} \sum_{t'=T_b}^{T_e} 
\LB\xv_{f', t'} - \Wm_{f,t} \yv_{f', t'} \RB
\LB\xv_{f', t'} - \Wm_{f,t} \yv_{f', t'} \RB\htp
 \RSB \nonumber .
\end{align}
}
\hspace{-3pt}Therefore, $\Wm_{f,t}$ nulls the gradient 

\vspace{-5pt}
{\small
\begin{align}
\nabla_{\Wm_{f,t}} \mathcal{L}(\Wm_{f,t})
 = & 2\Wm  \EE \LSB
\sum_{f'=F_b}^{F_e} \sum_{t'=T_b}^{T_e} 
 \yv_{f', t'} \yv_{f', t'}\htp  \RSB \\
&- 2 \EE \LSB
\sum_{f'=F_b}^{F_e} \sum_{f'=T_b}^{T_e} 
\xv_{f', t'} \yv_{f', t'}\htp \RSB
\stackrel{!}{=} 0 \nonumber
\end{align}
}
which leads to

\vspace{-5pt}
{\small
\begin{align}
&\Wm_{f,t} 
= \LB \EE \LSB
\sum_{f'=F_b}^{F_e} \sum_{t'=T_b}^{T_e} 
 \xv_{f',t'} \yv_{f',t'}\htp  \RSB \RB\\
 & \LB \EE \LSB
\sum_{f'=F_b}^{F_e} \sum_{t'=T_b}^{T_e} 
 \yv_{f',t'} \yv_{f',t'}\htp  \RSB \RB^{-1}\nonumber \\
&=\LB 
\sum_{f'=F_b}^{F_e} \sum_{t'=T_b}^{T_e} 
 \widehat{\Hm}_{f',t'}\htp  \RB\\
 &\LB 
\sum_{f'=F_b}^{F_e} \sum_{t'=T_b}^{T_e} \LB
 \widehat{\Hm}_{f',t'} \widehat{\Hm}_{f',t'}\htp  
 + \EE \LSB \widetilde{\Hm}_{f',t'} \widetilde{\Hm}_{f',t'}\htp \RSB
 + \sigma^2\Id_{N_m}
 \RB \RB^{-1}. \nonumber
\end{align}
}

\bibliographystyle{IEEEtran}
\bibliography{IEEEabrv, bib_abrv, bibliography}

\end{document}